\documentclass[12pt]{article}
\usepackage{type1cm,amsmath,amssymb,indentfirst}
\usepackage[dvips]{graphicx}
\usepackage{subfigure}

\newcommand{\6}{\partial}
\newcommand{\na}{\nabla}

\newcommand{\al}{\alpha}
\newcommand{\be}{\beta}
\newcommand{\ga}{\gamma}
\newcommand{\de}{\delta}

\newcommand{\ro}{\rho}

\newcommand{\m}{\mu}
\newcommand{\n}{\nu}

\newcommand{\et}{e_\theta}
\newcommand{\ep}{e_\phi}

\newcommand{\fn}{f_n(\theta)}
\newcommand{\A}{\mathcal{A}}
\newcommand{\J}{\mathcal{J}}

\newcommand{\Y}{\mathbb{Y}}
\newcommand{\proj}{\mathcal{P}}

\newcommand{\no}{\notag \\}

\newcommand{\lS}{\lambda^\text{S}_\kappa}
\newcommand{\lV}{\lambda^\text{V}_\kappa}
\newcommand{\lT}{\lambda^\text{T}_\kappa}
\newcommand{\CPn}{$\mathbb{CP}^{n-1}$}
\newcommand{\tY}{\mathbb{\tilde Y}}

\title{
Instability in near-horizon geometries of 
even-dimensional Myers-Perry black holes
}

\author{ Norihiro Tanahashi$^1$ and Keiju Murata$^{2,3}$
\\ 
{\small 
$^1$Department of Physics, University of California, One Shields Avenue, Davis, CA
95616, USA
}
\\
{\small
$^2$
DAMTP, Centre for Mathematical Sciences, Wilberforce Road, Cambridge CB3 0WA, UK
}
\\
{\small
$^3$
Yukawa Institute for Theoretical Physics, Kyoto university, Kyoto, 606-8502,
Japan
}
}

\begin{document}

\maketitle

\begin{abstract}
We study the gravitational, electromagnetic and scalar field perturbations
on the near-horizon geometries of
the even-dimensional extremal Myers-Perry black holes.
By dimensional reduction, 
the perturbation equations are reduced to effective equations of motion in AdS$_2$.
We find that some modes in the gravitational perturbations violate the
Breitenl\"ohner-Freedman bound in AdS$_2$. 
This result suggests that the even-dimensional (near-)extremal Myers-Perry
black holes are unstable against gravitational perturbations.
We also discuss implications of our results to the Kerr-CFT correspondence.
\end{abstract}

\section{Introduction}
\label{Sec:Intro}

Black hole solutions have been fundamental objects in the study of general
relativity theory. 
Higher-dimensional spacetimes introduced in some particle physics models
motivated to consider their generalizations to higher dimensions, and as a
result diverse solutions of black objects
were found~\cite{Emparan:2008zz,Emparan:2008eg}.
A feature specific to those higher-dimensional black hole solutions is that they
may be unstable, 
even though
the Schwarzschild and Kerr black holes in four dimensions are stable.
Those instabilities signal appearance of new blanches of black hole
solutions, and also tell us that some other solutions will be realized as a
result of time evolution.
As such, the stability analysis of the higher-dimensional black objects is
one of the most important steps 
to understand the phase structure of the solutions in the higher
dimensions and also the final states of the higher-dimensional spacetimes.

The generalizations of the (four-dimensional) Kerr black holes 
in $d$-dimensional spacetime 
are the Myers-Perry (MP) black holes~\cite{Myers:1986un}, which may have 
$\lfloor(D-1)/2\rfloor$ different angular momenta.
By an intuitive argument, 
it was predicted that ``ultraspinning'' MP black holes
should suffer from an instability~\cite{Emparan:2003sy}. 
Here, by ultraspinning, we mean that 
Hessian matrix, 
$H_{\alpha\beta}=-\partial^2S/\partial x^\alpha \partial x^b$
($x^\alpha=M,J_i$), 
for a black hole with mass $M$, angular momenta $J_i$
and horizon entropy $S$
has at least two negative eigenvalues~\cite{Dias:2010eu}.\footnote{
In Ref.~\cite{Dias:2010eu}, it was shown that any vacuum stationary black holes have at least one
negative mode in the Hesse matrix. 
It is a manifestation of the property that black holes have negative specific heat.
Ultraspinning MP black holes have at least one more negative mode
in addition to that.
}
%
%
Refs.~\cite{Kodama:2009bf,Dias:2009iu,Shibata:2010wz,Dias:2010maa} have clarified
the conditions for the ultraspinning instability
of the singly-spinning MP black holes,
and also Refs.~\cite{Dias:2010eu,Kunduri:2006qa,Murata:2008yx,Durkee:2010ea,Dias:2011jg}
studied the odd-dimensional MP black holes with equal angular momenta 
or all but one angular momenta are equal.
Since also the even-dimensional MP black holes with equal angular momenta
have ultra-spinning regime~\cite{Dias:2010eu},
we can expect the instability to occur even for these black holes.
The instability of 
these even-dimensional MP black holes 
is,
however, yet to be shown so far. 
A part of the reason is that the geometries of the even-dimensional ones
are cohomogeneity-2 and less symmetric compared to those of the odd-dimensional ones, 
which are cohomogeneity-1, and then the stability analysis becomes more involved.

One of our purposes is to fill this gap in our knowledge about the MP black holes by 
assessing the stability of the even-dimensional MP black holes with equal angular momenta.
Since the stability analysis of the full spacetime requires to solve 
complicated partial differential equations, it is desirable that we have more simpler ways to
obtain implications of the instability.
Ref.~\cite{Durkee:2010ea,Durkee:2010qu} initiated a study on a new analysis
method for such a demand.
Their method is based on a generalization of the Geroch-Held-Penrose formalism to higher
dimensions~\cite{Durkee:2010xq}.
They chose some combinations of the Weyl tensor components as perturbation
variables, and found that the variables satisfy decoupled equations of
motion when the near-horizon geometries of extreme black holes 
are taken as the background spacetimes.
Their method can be regarded as higher-dimensional generalization of
the Teukolsky formalism~\cite{Teukolsky:1972my},
although it is applicable only to the Kundt spacetimes.\footnote{
The Kundt spacetime is the spacetime admitting a null geodesic congruence with vanishing
expansion, rotation and shear.
All known near-horizon geometries are the Kundt spacetimes.
}
In Ref.~\cite{Durkee:2010ea}, it was conjectured that 
the instability of the near-horizon 
geometries imply the instability of the original full spacetime.
A proof for this statement is provided for the scalar field perturbations, and 
some pieces of evidence were found for the gravitational perturbations,
especially on the background of the odd-dimensional MP black holes.
In this paper, we assume this conjecture to be correct and study the
stability of the even-dimensional MP black holes with equal angular momenta 
by examining the perturbations on their near-horizon geometries.

We also have a motivation from the Kerr-CFT correspondence~\cite{Guica:2008mu},
which is a conjecture 
that near-extreme black holes are described by 2d CFTs. 
From the fall-off behavior of the perturbations on near-horizon geometries, 
we can read off the conformal weights of right sectors in the dual
CFTs~\cite{Durkee:2010ea,Murata:2011my,Dias:2009ex,Amsel:2009ev}.
In four and five dimensions, 
some interesting properties of the conformal weights have been found 
for the operators corresponding to the axisymmetric perturbations.
One of them is the integerness of the conformal weights:
all the conformal weights for vacuum near-horizon geometries take integral values.
Another one is the universality of the conformal weights:
all the $U(1)\times U(1)$ symmetric five-dimensional vacuum near-horizon 
geometries with vanishing cosmological constant
share the same sequences of conformal weights.\footnote{
In four dimensions, the near-horizon extreme Kerr 
geometry is the unique $U(1)$ symmetric vacuum 
near-horizon geometry~\cite{Kunduri:2008rs,Kunduri:2007vf} 
and does not have any dimensionless parameters.
Thus, the universality is trivial in four dimensions.
}
This result is surprising 
because these five-dimensional near-horizon geometries contain
dimensionless parameters
on which the conformal weights may depend in principle.
In addition to that, five-dimensional extreme black holes can have
$S^3$ or $S^2\times S^1$ horizon topology~\cite{Kunduri:2008rs,Hollands:2009ng}.
Nevertheless,
those various near-horizon geometries share the common sequences of the conformal weights.
This result suggests that there is a ``universal sector'' present in all the
CFTs dual to extreme rotating black holes in five dimensions.
These features have been observed only in four and five dimensions.
It was also observed that the effective mass of the perturbative fields 
becomes non-integer but rational numbers 
for any parameters when the 
background near-horizon geometry is unstable~\cite{Durkee:2010ea}.
We will examine if these properties persist 
even for the even-dimensional MP black holes
we are focusing on.

In our analysis method, which was used also in Ref.~\cite{Durkee:2010ea},
the perturbation equations of the near-horizon geometries are 
Kaluza-Klein reduced to an equation of motion of a massive charged scalar field
on AdS$_2$ spacetime whose effective mass is determined by the eigenvalues of
the angular part of the perturbation equations.
For the background of the even-dimensional MP black holes, those 
angular part equations are reduced to one-dimensional ordinary differential 
equations (ODEs) which are coupled to each other if we classify the perturbations
into scalar/vector/tensor modes according to their transformation properties
on \CPn contained in the spatial geometry on the horizon.
We solve those equations using the numerical technique employed in 
Ref.~\cite{Murata:2011my}, which analyzed the conformal weights for the general 
near-horizon geometries in five dimensions.

This paper is organized as follows.
After introducing the near-horizon geometries and the perturbation equations
in Sec.~\ref{Sec:Basics}, we show the results of the stability analysis for the 
gravitational perturbations in Sec.~\ref{gravinst}.
As a result, we find certain modes in these perturbations become unstable.
In Sec.~\ref{Scalar-EM}, we show the results for the scalar and 
electromagnetic field perturbations.
These fields are stable in the full geometries, and we observe that the 
analysis of the near-horizon geometries gives results consistent with
this fact.
As a byproduct of our analysis, we find that
the spectrum of the mass of the effective AdS$_2$ field, which is given by 
the eigenvalues of the angular part of the perturbations, show
a well-organized structure. We give comments on this property in 
Sec.~\ref{Sec:eigenvalues}.
We also discuss the implications of our results
to the Kerr-CFT correspondence in Sec.~\ref{Sec:Kerr-CFT},
especially about the integerness and universality of the conformal weights
mentioned above.
Sec.~\ref{Sec:Summary} is devoted to the summary and discussions, and 
appendices show the technical details of our analysis.

\section{Perturbations of near-horizon geometries}
\label{Sec:Basics}

We summarize the near-horizon geometries of the even-dimensional MP black 
holes with equal angular momenta, and also the perturbation equations on this 
background in this section.
We also introduce the instability condition for the near-horizon geometries 
which is argued in Ref.~\cite{Durkee:2010ea}.

\subsection{Near-horizon geometry}

In our study, we focus on the extremal MP black holes 
in $d=2n+2$ ($n\geq 2$) dimensions with all the angular momenta set equal.
Their near-horizon geometries can be written as a fibration over
AdS$_2$ given by~\cite{Figueras:2008qh}
\begin{equation}
ds^2 
=
L(\theta)^2\left(-r^2 dt^2 + \frac{dr^2}{r^2}\right)
+ 
\Theta^2(\theta)
d\theta^2
+ 
\Psi^2(\theta)
\hat g_{\alpha\beta} dx^\alpha dx^\beta
+ 
\Phi^2(\theta)
\left(d\phi + \A - \Omega rdt\right)^2
,
\label{metric}
\end{equation}
where $0\leq\theta\leq\pi$, 
\begin{equation}
\begin{gathered}
L^2(\theta)=\frac{a^2\fn}{(2n-1)^2},\qquad
\Theta^2(\theta) = \frac{a^2\fn}{2n-1},
\qquad
\Psi^2(\theta) = \frac{2na^2\sin^2\theta}{2n-1},
\qquad
\Phi^2(\theta) = \frac{4n^2a^2\sin^2\theta}{(2n-1)\fn},\\
\fn=1+(2n-1)\cos^2\theta,
\qquad
k^\phi = -\frac{1}{n\sqrt{2n-1}}\equiv\Omega,
\end{gathered}
\end{equation}
$\phi$ is $2\pi$-periodic
and 
$\hat g_{\alpha\beta} dx^\alpha dx^\beta$ ($\alpha,\beta = 1,\ldots,2n-2$)
is the Fubini-Study metric on \CPn
with K\"ahler form $\J=\frac12 d\A$
normalized to have 
$R_{\alpha\beta}=2n\hat g_{\alpha\beta}$.
From the near-horizon metric, we can read off 
the spatial metric on the horizon (the surface of constant $t$ and $r$) as
\begin{equation}
d s^2 
= 
\Theta^2(\theta) d\theta^2 
+ \Psi^2(\theta) \hat g_{\alpha\beta} dx^\alpha dx^\beta
+ \Phi^2(\theta) \left(d\phi + \A \right)^2.
\label{horizonmetric}
\end{equation}

\subsection{Perturbation equations}

Ref.~\cite{Durkee:2010ea} gave a prescription to obtain separated perturbation 
equations for general near-horizon geometries. 
We explain it briefly below.

To describe the perturbations of the near-horizon geometry, we introduce
a null basis $\{\ell,n,m_i\}$, where 
$\ell= L(\theta)(-rdt + dr/r)/\sqrt2$, $n= L(\theta)(rdt + dr/r)/\sqrt2$
and $m_i$ are orthonormal spacelike vectors orthogonal to both $\ell$ and $n$.
We define $\Omega_{ij}\equiv C_{abcd}\ell^a m_i^b\ell^c m^d_j$
and $\varphi_i\equiv F_{ab}\ell^a m^b_i$,
where $C_{abcd}$ is the Weyl tensor and $F_{ab}$ is the electromagnetic field strength,
and use them as the perturbation variables.
As for the scalar field satisfying $\bigl(\square-M^2\bigr)\psi=0$, we use the scalar 
field itself as the perturbation variable.
In this paper, we focus only on axisymmetric modes along $\phi$
direction: $\partial_\phi\Omega_{ij}=\partial_\phi \varphi_i=\partial_\phi \psi=0$.
For these perturbations, we can take the separation ansatz as
\begin{equation}
\psi = \chi_0(t,r)Y(\theta,x^\alpha), \qquad
\varphi_i = \text{Re}\left[\chi_1(t,r)Y_i(\theta,x^\alpha)\right], \qquad
\Omega_{ij} = \text{Re}\left[ \chi_2(t,r)Y_{ij}(\theta,x^\alpha)\right].
\end{equation}
Then, the perturbation equation is separated as follows. 
The radial equations are given by massive charged Klein-Gordon 
equations in AdS$_2$ with homogeneous electric fields
\begin{equation}
\left[ \left( \nabla_2 - i q_s A_2 \right)^2 - q_s^2 - \lambda_s \right]
\chi_s = 0,
\quad
\left(s=0,1,2\right)
\label{radialequations}
\end{equation}
where $\nabla_2$ is the covariant derivative on AdS$_2$ 
$\bigl(ds^2 = -r^2 dt^2 +\frac{dr^2}{r^2}\bigr)$, $A_2=-rdt$ is the effective gauge field
which appeared as a result of the dimensional reduction,
and $q_s \equiv is$ are the effective $U(1)$ charges.
The angular equations are given by
\begin{equation}
\mathcal{O}^{(0)}Y = \lambda_0 Y,
\qquad
\bigl( \mathcal{O}^{(1)}Y \bigr)_\mu = \lambda_1 Y_\mu,
\qquad
\bigl( \mathcal{O}^{(2)}Y \bigr)_{\mu\nu} = \lambda_2 Y_{\mu\nu},
\label{angularequations}
\end{equation}
where the operators $\mathcal{O}^{(s)}$ are defined as
\begin{align}
\mathcal{O}^{(0)} Y &= 
-\na_\mu \left( L^2 \na^\mu Y\right)
+L^2 M^2 Y,
\label{scalar}
\\
\bigl(\mathcal{O}^{(1)}Y\bigr)_\mu
&= 
- \frac{1}{L^2} \na^\ro \left( L^4 \na_\ro Y_\mu\right)
+\left(
2-\frac{5}{4L^2}k_\nu k^\nu 
\right)Y_\mu
+ L^2 \left(R_{\mu\nu} + \frac12 R g_{\mu\nu}\right)Y^\nu
\no
& \qquad\qquad\qquad\qquad ~~~\,
+ \left(
-\frac12 \left(dk\right)_{\mu\nu} + 2\left(k-d\bigl(L^2\bigr)\right){\!}_{[\mu}
\na_{\nu]} -\frac{1}{L^2}\left(dL^2\right){\!}_{[\mu} k_{\nu]}
\right)Y^\nu,
\label{Maxwell}
\\
\bigl(\mathcal{O}^{(2)}Y\bigr)_{\m\n}
&= 
-\frac{1}{L^4}\na^\ro\left(L^6 \na_\ro Y_{\m\n}\right)
+\left( 6-\frac{4}{L^2}k_\ro k^\ro
\right) Y_{\m\n}
+ 2L^2\left(R_{(\m|\ro}+R  g_{(\m|\ro}\right) Y^\ro{\!\!\!}_{~|\n)}
- 2L^2 R_{\m~\n}^{~\,\ro~\sigma}Y_{\ro\sigma}
\no
&\quad\,
+ \left[
-(dk)_{(\m|\ro} 
- \frac{2}{L^2}\left(d(L^2)\wedge k\right){\!}_{(\m|\ro}
+2\left(k-d(L^2)\right){\!}_{(\m|}\na_\ro
- 2\left(k-d(L^2)\right){\!}_\ro \na_{(\m|}
\right] Y^\ro{\!\!\!}_{~|\n)}~,
\label{gravity}
\end{align}
where the covariant derivative 
$\nabla_\mu$ and the curvature tensors such as $R_{\mu\nu\rho\sigma}$ are 
defined with respect to the spatial metric on the horizon~(\ref{horizonmetric}).

The solution of Eq.~(\ref{radialequations}) at large $r$ behaves as 
$\chi_s\sim r^{-\Delta_\pm}$, where
\begin{equation}
\Delta_\pm = \frac12 \pm \sqrt{\frac14 +\lambda_s }~.
\label{conformalweight}
\end{equation}
Following the arguments of Ref.~\cite{Durkee:2010ea}, we shall call the near-horizon 
geometry to be unstable when $\lambda_s$ violates the effective 
Breitenl\"ohner-Freedman (BF) bound, i.e., $\lambda_s < -1/4$,
and $\Delta_\pm$ becomes complex.
Ref.~\cite{Durkee:2010ea} conjectured that the sufficient condition for the 
full black hole geometry to be unstable is that the near-horizon geometry is 
unstable against axisymmetric perturbations.
This conjecture is proved in the case of the scalar field, 
and it was suggested that a similar argument applies also to the gravitational
perturbations.

In this paper, we argue the stability of the even-dimensional MP black 
holes with equal angular momenta based on this conjecture.
We solve Eqs.~(\ref{angularequations}) 
analytically if possible and numerically otherwise to obtain
$\lambda_s$. 
Applying the conjecture to the resultant $\lambda_s$,
we will argue the stability of the full black hole geometries.


\section{Instability against gravitational perturbations}
\label{gravinst}

We start our discussions
from the gravitational perturbations on the near-horizon geometries.
The gravitational perturbations can be decomposed by tensor, vector and
scalar harmonics on the
base space \CPn, which are labeled by a principal quantum number
$\kappa=0,1,2,\ldots$~.
By the mode decompositions, the eigenvalue equations~(\ref{gravity})
reduce to coupled ODEs which depend only on
the $\theta$ coordinate.
Solving the ODEs, we obtain the eigenvalues for the operator
$\mathcal{O}^{(2)}$. The computation is straightforward but too technical
to show the details here.
Hence, we describe only some important results in this section
and we defer the details of the calculations to the appendices.

\subsection{Gravitational tensor modes}

We start from the simplest components, which are the gravitational
tensor modes.
The tensor modes are decomposed by tensor harmonics in \CPn,
and as a result
the eigenvalue equations~(\ref{gravity})
reduce to an ODE for a single coordinate $\theta$ given by
Eq.~(\ref{eq_gravtensor}).
For $n=2$, the tensor harmonics do not exist on $\mathbb{CP}^1=S^2$.
For $n\geq 3$, we obtain an analytical expression for the
eigenvalues of $\mathcal{O}^{(2)}$ as
\begin{equation}
 \lambda_2 = 
\frac{
2\left\{ \kappa+n(\ell+1)-1 \right\}\left\{ \kappa + n(\ell+1) \right\}
}{n(2n-1)}
-\ell(\ell+1)
-\frac{2(n^2-3n+1)}{n(2n-1)} -\frac{2\sigma}{n},
\qquad
(\ell=0,1,2,\ldots)
\label{lambda_gravtensor}
\end{equation}
where $\kappa=0,1,2,\ldots$ is the principal quantum number which labels
the tensor harmonics. The parameter $\sigma=\mp 1$ separates the tensor
harmonics into Hermitian and anti-Hermitian ones. 
See App.~\ref{App:gravtensor} for the further details of the tensor harmonics.
Another integer $\ell$ is the quantum number along $\theta$ direction,
which parametrizes the $\theta$ dependence of $Y_{\mu\nu}$.
This $\lambda_2$ is always non-negative, and does not violate the BF bound.
Thus, there is no indication of instability in the tensor modes.


\subsection{Gravitational vector modes}

Next, we study the gravitational vector modes.
The vector modes are decomposed by vector harmonics in \CPn and their derivatives.
For $n=2$, the vector harmonics do not exist on $\mathbb{CP}^1=S^2$.
For $n\geq 3$, we find the eigenvalues are given by
\begin{equation}
 \lambda_2 = 
\frac{
2\left\{
\kappa+n\left(\ell+1\right)
\right\}
\left\{
\kappa + n\left(\ell+1\right)+2
\right\}
}{
n\left(2n-1\right)
}
-\ell\left(\ell+1\right)
+\frac{C}{2n-1},
\label{lambda_gravVector}
\end{equation}
where $\kappa=0,1,2,\ldots$, $\ell = \ell_0,\,\ell_0+1,\ldots$
and $(C, \ell_0)$ are the set of integers given by
\begin{equation}
 \left(C, \ell_0\right) =
\left( 4-2n, 0 \right), \quad
\left( 4, -1 \right), \quad
\left( 4, 0 \right), \quad
\left( 4, +1 \right).
\label{lambda_gravVector_list}
\end{equation}

The gravitational vector modes are described by four free variables,
as we see in App.~\ref{App:gravvector}.
This is why we have four kinds of the eigenvalues 
described by different integer sets $(C, \ell_0)$
as listed in Eq.~(\ref{lambda_gravVector_list}). 
The eigenvalues are all positive, and hence no instability is implied.

\subsection{Gravitational scalar modes}

Finally, we consider the gravitational scalar modes.
We can expand the scalar modes by scalar harmonics in \CPn and their
derivatives.
As a result of mode decomposition,
we obtain the coupled ODEs of ten variables
in general cases ($\kappa>1$ and $n>2$) 
as the eigenvalue equations to solve.
In the cases of $\kappa=0,1$ and/or $n=2$, 
some of the unknown variables drop
out and we have to treat those cases separately.
In any case, we find that 
the eigenvalues are written in an unified expression as 
\begin{equation}
\lambda_2 =
\frac{
2 
\left\{ \kappa + n \left(\ell + 1\right) - 1 \right\}
\left\{ \kappa + n \left(\ell + 1\right)  \right\}
}{n(2n-1)} - \ell(\ell+1)
+ \frac{C}{2n-1},
\label{lambda_gravScalar_simple}
\end{equation}
where $\kappa=0,1,2,\ldots$ and $\ell = \ell_0,\ell_0+1,\ldots$~.
For a general case with $\kappa>1$ and $n>2$, we have ten eigenvalues described
by the integer sets $(C,\ell_0)$ given by
\begin{multline}
 (C,\ell_0) = 
(-2 (n - 1), 0 ), \quad
 (0, -1),\quad
(0, 0),\quad
(0, +1),
\\
(2 n, -2),\quad
(2 n, -1),\quad
(2 n, 0),\quad
(2 n, 0),\quad
(2 n, +1),\quad
(2 n, +2).
\label{lambda_gravScalar_simple2}
\end{multline}
Note that the constant $C$ takes only three different values,
that is, $C= -2(n-1)$, $0$ and $2n$, and 
each of them is associated with
one, three and six eigenvalues labeled by different $l_0$, respectively.
For the special values of $\kappa$ and $n$ mentioned above,
the eigenvalues are described by
\begin{align}
\kappa=0:&&
(C,\ell_0) &=
(-2 (n - 1), +2),
\quad
(0, +2), 
\quad 
(2 n, +2)
\label{lambda_gravscalar_list_k0}
\\
\kappa=1,n=2:&&
(C,\ell_0) &= 
(-2 (n - 1), 0),~~
(0, 0),~~
(0, 0),~~
(0, +1),~~
(2 n, 0),~~
(2 n, +1),~~
(2 n, +2)
\label{lambda_gravscalar_list_k1n2}
\\
\kappa=1,n>2:&&
(C,\ell_0) &= 
 (-2 (n - 1), 0), \quad
(0, 0),\quad
(0, 0),\quad
(0, +1),
\notag \\
&&& \hspace{5.5cm}
(2 n, 0),\quad
(2 n, 0),\quad
(2 n, +1),\quad
(2 n, +2)
\label{lambda_gravscalar_list_k1}
\\
\kappa>1,n=2:&&
(C,\ell_0) &= 
 (-2 (n - 1), 0), \quad
 (0, -1),\quad
(0, 0),\quad
(0, +1),
 \notag\\
&& &\hspace{3.09cm}
(2 n, -2),\quad
(2 n, -1),\quad
(2 n, 0),\quad
(2 n, +1),\quad
(2 n, +2).
\label{lambda_gravscalar_list_n2}
\end{align}
Comparing
Eqs.~(\ref{lambda_gravscalar_list_k0})--(\ref{lambda_gravscalar_list_n2})
to Eq.~(\ref{lambda_gravScalar_simple2}), we notice that 
some of the eigenvalue sequences described by the integers listed in 
Eq.~(\ref{lambda_gravScalar_simple2})
drop out in the special cases with low $\kappa$ and $n$.
We notice also that $l_0$ increases in some of the remaining sequences,
which implies that low $\ell$ modes existing in general cases
are truncated in the special cases.

From the expression of the eigenvalues, 
we can find the parameter region for the instability, that is,
$\lambda_2<-1/4$.
The eigenvalues other than $(C, \ell_0)=(2n, -2)$ in $\kappa>1$ and
$n\geq 2$ 
are shown to be non-negative, 
and thus they do not imply instability.
Now, let us examine the eigenvalues for $(C, \ell_0)=(2n, -2)$.
For $\ell\geq -1$, we can show that the eigenvalues are non-negative
again. However, for the lowest mode $\ell=-2$, we have
\begin{equation}
\lambda_2 = \frac{2(\kappa-1)(\kappa-2n)}{n(2n-1)}.
\end{equation}
This eigenvalue can be smaller than $-1/4$ depending on $\kappa$ and
$n$. The instability condition can be written as
\begin{equation}
\kappa_- < \kappa < \kappa_+~,
\qquad
\kappa_\pm \equiv 
\frac{ 4 n+2 \pm \sqrt{2(2n-1)(3n-2)} }{4}.
\label{k_pm}
\end{equation}
We show the values of $\kappa_\pm$ for some $n \geq 2$ ($d=2n+2\geq 6$) 
in Table~\ref{Table:k_pm}.
We find that a several numbers of 
$\kappa$ satisfy the instability condition for any 
$n\geq 2$.
We find also that the parameter region for the instability widens
as we increase the number of dimensions $n$.
It would be interesting to search for instability in the full spacetime 
geometry corresponding to these unstable modes on the near-horizon geometries.
\begin{table}[htbp]
\centering
\begin{tabular}{|c||c|c|c|c|c|c|c|c|} \hline
$d$ & 6 & 8 & 10 & 12 & 14 & 16 & 18 & 20  \\ \hline 
$\kappa_-$ & 1.28  & 1.41  & 1.54  & 1.68  & 1.81  & 1.94  & 2.08  & 2.21  \\ \hline
$\kappa_+$ & 3.72  & 5.59  & 7.46  & 9.32  & 11.19 & 13.06 & 14.92 & 16.79 \\ \hline
\end{tabular}
\caption{
$\kappa_\pm$ against the number of dimensions $d$. 
The near-horizon geometries are unstable against gravitational 
scalar perturbation satisfying 
$\kappa_- < \kappa < \kappa_+$.
}
\label{Table:k_pm}
\end{table}

\section{Scalar and electromagnetic field perturbations}
\label{Scalar-EM}

Having finished the study on the gravitational perturbations, we move on to
the scalar and electromagnetic field perturbations.
These perturbations are guaranteed to be stable 
on the full geometry by the following argument.
These fields satisfy dominant energy conditions
and, in addition to that, there is a global timelike Killing vector in 
the full geometry. Then, 
we can construct an energy integral whose integrand is non-negative 
everywhere~\cite{Hawking:1999dp,Kodama:2007sf}.
Hence, any instability cannot occur.
Therefore, 
if we show the stability of the near-horizon geometry
against these perturbations, 
we can give a non-trivial check of 
the conjecture in Ref.~\cite{Durkee:2010ea}.

As for the scalar field perturbations, we may take the scalar field as the
perturbation variable, and it is decomposed by the scalar harmonics on
\CPn. We show the result for this case in Sec.~\ref{Sec:scalar}.
As for the electromagnetic field perturbations, the perturbation variables are
decomposed into the vector and scalar modes on \CPn. We show the results for 
each mode in Sec.~\ref{Sec:EMvector} and Sec.~\ref{Sec:EMscalar}, respectively.
We will see that any perturbations are stable, which is consistent with the
conjecture mentioned above.

\subsection{Scalar field perturbations}
\label{Sec:scalar}

The scalar field can be decomposed by scalar harmonics in \CPn.
As a result, we obtain a single ODE 
as the perturbation equation, which is given by Eq.~(\ref{O0}).
When the scalar field is massless ($M=0$), 
we analytically find the eigenvalues to be given by
\begin{equation}
\lambda_0 
= 
\frac{
2
\left\{
\kappa + n(\ell+1) -1
\right\}
\left\{
\kappa+n(\ell + 1)
\right\}
}{n(2n-1)}
- \ell(\ell+1) - \frac{2(n-1)}{2n-1},
\label{lambda0}
\end{equation}
where $\kappa=0,1,2,\ldots$ and $\ell = 0,1,2\ldots$~.
This $\lambda_0$ is non-negative for any 
$\kappa$, $\ell$ and $n\geq 2$,
and thus the AdS$_2$ BF bound ($\lambda_0\geq-1/4$) is not
violated.
This is consistent with the stability of the scalar field perturbation on
the full geometry.
We found evidence that this property persists even for a massive scalar field ($M>0$).
See Sec.~\ref{App:scalar} for details.

\subsection{Electromagnetic vector modes}
\label{Sec:EMvector}

Electromagnetic fields can be decomposed by vector and scalar harmonics
in \CPn.
Here, we consider the vector modes.
For $n=2$, there is no vector harmonics on $\mathbb{CP}^1$.
For $n>2$, we analytically find 
the eigenvalues for the vector modes are given by
\begin{equation}
 \lambda_1 
=
\frac{
2 \left\{ \kappa + n (\ell+1) \right\}
\left\{
\kappa + n(\ell + 1) + 2
\right\}
}{n(2n-1)}
- \ell(\ell+1)
+ \frac{4-2n}{2n-1},
\label{lambda_EVvector}
\end{equation}
where $\ell=0,1,2,\ldots$~.
These eigenvalues are always positive, and thus these modes are expected
to be stable.

\subsection{Electromagnetic scalar modes}
\label{Sec:EMscalar}

Finally, we consider the electromagnetic scalar modes, which can be
decomposed by scalar harmonics on \CPn and their derivatives.
As a result of the mode decomposition,
we obtain coupled ODEs for four variables
in a general case ($\kappa>1$) 
as the eigenvalue equations to solve.
When $\kappa=0$, the scalar harmonics becomes a constant on 
\CPn and the scalar-derived
vectors vanish. 
Then, we have only two variables to solve.
For both cases of $\kappa\geq 1$ and $\kappa=0$, 
the eigenvalues are written by an unified expression as
\begin{equation}
 \lambda_1 =
\frac{
2\left\{ \kappa + n(\ell+1) - 1 \right\}
\left\{ \kappa + n(\ell+1) \right\}
}{n(2n-1)} 
- \ell(\ell+1)
+ \frac{C}{2n-1},
\label{lambda_EMscalar}
\end{equation}
where $\ell = \ell_0, \ell_0+1, \ldots$~.
For  $\kappa>0$, the integer sets $(C,\ell_0)$ given by
\begin{equation}
 (C, \ell_0) = 
(-2(n-1),0), \quad
(0,-1), \quad
(0,0), \quad
(0,+1).
\label{lambda_EMscalar_list}
\end{equation}
For  $\kappa=0$, they reduce to
\begin{equation}
 (C,\ell_0) = 
(-2(n-1), +1), \quad
(0, +1).
\label{lambda_EMscalar_list_k0}
\end{equation}
The eigenvalues for these parameter sets are non-negative, and instability
is not implied. This result along with that in the previous section
are consistent with the stability of the full
geometry against the electromagnetic perturbations.

\section{
Comments on the effective mass spectrum
}
\label{Sec:eigenvalues}

In this section,
we give brief comments on the spectrum of the effective mass of 
the radial field on AdS$_2$, namely,
the relationships among the eigenvalues 
we have clarified in the previous sections.

Comparing Eq.~(\ref{lambda_gravVector_list}) to 
Eq.~(\ref{lambda_EVvector}), we notice that 
the eigenvalues of the gravitational vector modes
corresponding to the first one in 
Eq.~(\ref{lambda_gravVector_list}),
$(C,\ell_0)=(4-2n,0)$,
coincide exactly with 
those of the electromagnetic vector modes given by Eq.~(\ref{lambda_EVvector}).
Since the gravitational vector modes have also the eigenvalues with
$C=4$, the eigenvalue sets of the electromagnetic vector modes is completely
covered by those for the gravitational vector modes.

We also find similar relationships among 
the eigenvalues of the scalar perturbations of the
various fields as follows.
\begin{itemize}
 \item Scalar field and electromagnetic scalar modes:

Comparing Eq.~(\ref{lambda_EMscalar}) with Eq.~(\ref{lambda0}), 
we find that the eigenvalues of the electromagnetic scalar modes coincide with the 
those of the scalar field perturbations if we set 
$C=-2(n-1)$. 
The first ones in Eqs.~(\ref{lambda_EMscalar_list}) and
(\ref{lambda_EMscalar_list_k0})
for the electromagnetic scalar modes
share such a property, while the eigenvalue for $(C,\ell)=(-2(n-1),0)$ do not appear
in the electromagnetic scalar modes with $\kappa=0$. 
It means that the eigenvalues of the scalar field and electromagnetic scalar mode
perturbations for $C=-2(n-1)$ 
are almost identical to each other for any $\kappa$ and $n$,
and the only exception is the eigenvalues for $\ell=0$ 
of the electromagnetic scalar modes with $\kappa=0$.

\item Electromagnetic and gravitational scalar modes:

The eigenvalues corresponding to the first four integer sets in
Eq.~(\ref{lambda_gravScalar_simple2}) for the gravitational scalar modes,
which have $C=0$ and $C=-2(n-1)$,
coincide exactly with those of the 
electromagnetic scalar modes given by Eqs.~(\ref{lambda_EMscalar}) and 
(\ref{lambda_EMscalar_list}). 
When $\kappa=0,1$ and/or $n=2$,
we find some of the eigenvalues of the electromagnetic scalar modes,
especially those for low $\ell$ modes,
are missing in the eigenvalues with $C=0$ and $C=-2(n-1)$ of
the gravitational scalar modes,
as we can see from
Eqs.~(\ref{lambda_gravscalar_list_k0})--(\ref{lambda_gravscalar_list_n2}).
\end{itemize}

To summarize, we find the inclusion relations among the eigenvalues 
for the vector perturbations expressed as
\begin{equation}
\left\{ \lambda_1 \text{~of EM vector with~}C=4-2n\right\}
=
\left\{ \lambda_2 \text{~of grav.~vector with~}C=4-2n \right\},
\end{equation}
\begin{equation}
\left\{ \lambda_1 \text{~of EM vector} \right\}
\subset 
\left\{ \lambda_2 \text{~of grav.~vector} \right\}.
\end{equation}
Similarly, 
if we do not care about 
a small numbers of 
the exceptions for small $\kappa$ and $n$
mentioned above,
we have the relations among the eigenvalues 
for the scalar perturbations given by
\begin{equation}
\begin{aligned}
\left\{ \lambda_0 \text{~with~}C=-2(n-1)\right\}
&=
\left\{ \lambda_1 \text{~of EM scalar with~}C=-2(n-1)\right\}
\\
&=
\left\{ \lambda_2 \text{~of grav.~scalar with~}C=-2(n-1) \right\},
\end{aligned}
\label{lambda_top}
\end{equation}
\begin{equation}
\left\{ \lambda_1 \text{~of EM scalar with~}C=0\right\}
=
\left\{ \lambda_2 \text{~of grav.~scalar with~}C=0 \right\},
\label{lambda_second}
\end{equation}
\begin{equation}
\left\{ \lambda_0 \right\}
\subset
\left\{ \lambda_1 \text{~of EM scalar} \right\}
\subset 
\left\{ \lambda_2 \text{~of grav.~scalar} \right\}.
\label{lambda_total}
\end{equation}
If we care about the missing eigenvalues for $\kappa=0,1$ and/or $n=2$,
the equality in Eqs.~(\ref{lambda_top}) and (\ref{lambda_second}) should be
changed into ``$\supset$'', and Eq.~(\ref{lambda_total}) needs to be modified accordingly.

We also notice that the set of the eigenvalues for fixed $C$ have simple
organizations which are described by $\ell_0$ being integers around zero.
This property is most noticeable in the eigenvalue
list~(\ref{lambda_gravScalar_simple2}) for the gravitational scalar modes.
There is only one eigenvalue with $C=-2(n-1)$, which is associated with 
$\ell_0=0$.
As for the eigenvalues which share $C=0$, there are three of them, 
and they have $\ell_0=-1$, $0$ and $+1$. 
Similarly, there are six eigenvalues which share
$C=2n$, and they have $\ell_0=-2$, $-1$, $0$, $0$, $+1$ and $+2$. 
The eigenvalue sets for the scalar field and electromagnetic scalar modes
inherit this structure of the eigenvalues,
and we can see a similar relationship to hold
also between gravitational and
electromagnetic vector modes.

Similar relations for the eigenvalues can be seen also in 
the results of
Ref.~\cite{Durkee:2010ea,Murata:2011my}. 
The eigenvalues for various near-horizon geometries in five dimensions are
studied in Ref.~\cite{Murata:2011my}, and it was observed that the eigenvalues
of the scalar field perturbations are included in those for the electromagnetic
perturbations, and those for the gravitational perturbations cover both of them.
Ref.~\cite{Durkee:2010ea} studied the eigenvalues for the odd-dimensional MP
black holes with equal angular momenta.
By rewriting their results in the form similar to
Eq.~(\ref{lambda_gravScalar_simple}),
we find the the eigenvalues of the vector perturbations in that case is
expressed by a unified formula
\begin{equation}
 \lambda = \frac{
2
\left\{ \kappa + n\bigl(C' + \frac12\bigr) \right\}
\left\{ \kappa + n\bigl(C' + \frac12\bigr) + 2 \right\}
}{n(n-1)}
+ \frac{C}{2(n-1)},
\end{equation}
where $n\geq 2$ in this formula corresponds to $N+1$ in Ref.~\cite{Durkee:2010ea}.
The eigenvalues of the electromagnetic vector modes are expressed by 
$(C, C') = (-(n-4),0)$, while
the eigenvalues for the gravitational vector modes are expressed by
\begin{equation}
 (C,C') = 
\bigl(-(n-4),0\bigr),  \qquad
\left(4,-\frac12\right), \qquad
\left(4,+\frac12\right).
\end{equation}
The inclusion relation between those two modes and also 
the simple composition of the eigenvalues for $C=4$ resemble to those in our
case to some extent, that is, 
the eigenvalues for the electromagnetic vector modes appears as a part of the
eigenvalues for the gravitational vector modes, and the eigenvalues with $C=4$
in the latter are described by some numbers $C'$ around zero which defer by one
from each other.
Similarly, the eigenvalues for the scalar field, electromagnetic and
gravitational scalar modes are expressed by a single formula 
\begin{equation}
 \lambda = 
\frac{
2
\left\{ \kappa + n\left( C' + \frac12 \right) - 1 \right\}
\left\{ \kappa + n\left( C' + \frac12 \right) \right\}
}{n(n-1)}
+ \frac{C}{2(n-1)},
\end{equation}
where the constants $(C,C')$ are given by
\begin{align}
 \text{Scalar field:} &&
(C,C') &= \bigl( -(n-2), 0\bigr)
\\
 \text{EM scalar:} &&
(C,C') &= 
\bigl( -(n-2), 0\bigr), \quad
\left( 0, -\frac12 \right), \quad
\left( 0, +\frac12 \right)
\\
\text{Grav.~scalar:} &&
(C,C') &= 
\bigl( -(n-2), 0\bigr), \quad
\left( 0, -\frac12 \right), \quad
\left( 0, +\frac12 \right),
\notag \\
&&&
\quad ~
\bigl( -(n-2), -1 \bigr), \quad
\bigl( -(n-2), +1 \bigr), \quad
\left(3n-2, 0\right).
\label{odd_gravscalar}
\end{align}
It is fair to say that
the inclusion relations for the eigenvalues with $C=-(n-2)$ and $0$, and also
the composition of the eigenvalues with $C=0$ are similar to those in our cases.
The only exceptions are 
the eigenvalues which appears only for the gravitational scalar modes,
the last three sets in Eq.~(\ref{odd_gravscalar}). They do not share the same 
constant $C$, while the counterparts in our cases (the last six eigenvalues
in Eq.~(\ref{lambda_gravScalar_simple2}) with $C=2$) did.
Instead of that, the first, fourth and fifth eigenvalues in
Eq.~(\ref{odd_gravscalar}) shares the same $C=-(n-2)$ in the case of
the odd-dimensional MP black holes.

The origin for the composition of the eigenvalues described above is unclear so
far, though their simplicity tempts us to, naively thinking,
suspect the existence of hidden background mechanism to generate it.
It would be interesting to study the mathematical 
and physical origins of these properties studying, e.g.,
the perturbations of other near-horizon geometries.

\section{Conformal weight in the Kerr/CFT correspondence}
\label{Sec:Kerr-CFT}

In any four- and five-dimensional vacuum near-horizon geometry, 
operators dual to the
gravitational, electromagnetic and massless scalar field perturbations
preserving rotational symmetry 
have integer conformal
weights~\cite{Durkee:2010ea,Murata:2011my,Dias:2009ex,Amsel:2009ev}.
Let us study if such a property persists
in the case of the even-dimensional MP black holes we have studied.

As we have seen in Sec.~\ref{gravinst},
the near-horizon geometries have unstable modes 
in the gravitational scalar perturbations.
Conformal weights become complex for the unstable modes, 
as we can see from Eq.~(\ref{conformalweight}).
From the formulae for the eigenvalues we have shown in 
Secs.~\ref{gravinst} and \ref{Scalar-EM}, we find that the 
conformal weights for the most of stable modes take irrational numbers.
For example, 
a stable gravitational scalar modes for
$\kappa=0$, $n=2$ and $(C,\ell_0)=(2n,2)$
have a sequence of conformal weights given by
$h_R-1/2=\sqrt{201}/6$, $\sqrt{33}/2$, $\sqrt{417}/6,\ldots$~.
This result suggests the following scenarios
about the integerness of the conformal weights:
the conformal weights become integers only in four and five dimensions,
or they may take integer values 
even in higher dimensions
while they do so only when the background black holes are stable.

Below, we discuss another property of the conformal weights, 
the universality, mentioned in Sec.~\ref{Sec:Intro}.
In any five-dimensional vacuum near-horizon geometry, 
operators dual to axisymmetric perturbations have universal 
conformal weights. That is, they do not depend on parameters 
nor horizon topologies of extreme black holes.
We consider six dimensions for simplicity and check 
if the universality holds in this case.

To compare with the six-dimensional MP black holes, 
we consider the MP black string solution whose metric is given
by $ds^2=ds^2(\textrm{MP}_5)+dz^2$ where $ds^2(\textrm{MP}_5)$
is  the five-dimensional MP metric.
By the dimensional reduction along $z$-direction,
the gravitational perturbation on the six-dimensional MP black
string solution is decomposed into five-dimensional 
massless scalar, electromagnetic and gravitational perturbations
on the background of $ds^2(\textrm{MP}_5)$.
The conformal weights for those fields have been studied in 
Refs.~\cite{Durkee:2010ea,Murata:2011my}, 
and it was found that all the eigenvalues for these perturbations are integers.
On the other hand, the gravitational perturbations on the
near-horizon geometries of six-dimensional MP black holes has
rational but non-integer eigenvalues as we have seen in Sec.~\ref{gravinst}.
This result suggests again that the universality is a property 
exists only in four and five dimensions or only for stable black holes.

\section{Summary and discussion}
\label{Sec:Summary}

We studied the gravitational, electromagnetic and scalar field perturbations
on the near-horizon geometries of the even-dimensional extreme MP black
holes with equal angular momenta.
As a result of our study, we find that some modes in the gravitational
scalar modes become unstable while all the others do not.
This result implies that the instability for the corresponding modes
in the full background geometries, assuming the claim of
Ref.~\cite{Durkee:2010ea} to be correct. 
%
%
Ref.~\cite{Dias:2011jg} raised a question if there is any (near-)extremal
Myers-Perry black hole which is stable in $d \geq 6$, and our results
suggest that there is no such in even dimensions if the angular
momenta are set equal.
%
%
%
If the black holes are unstable in the extreme limit, 
it is reasonable to consider that they are also unstable 
sufficiently near extremality. 
It is important to confirm this expectation analyzing the stability of
the full geometry.

Recently, a new method for demonstrating dynamical instability
was established~\cite{Figueras:2011he,Hollands:2012sf}.
They found an inequality whose violation implies
black hole instability. 
The method based on this inequality
makes the stability analysis dramatically easier, 
because
the inequality can be evaluated only from 
initial data that describes a small perturbation
of the black hole
 and we do not need to solve time evolution from that data.
This method is a hopeful approach for demonstrating
the instability of the full geometries of the
even-dimensional MP black holes with equal angular momenta.

For small angular momenta, the MP black holes are
stable since higher-dimensional Schwarzschild black hoes are
stable~\cite{Ishibashi:2003ap,Konoplya:2007jv}.
Thus, at critical values of angular momenta, 
the stability changes and there should be a static perturbation.
The static perturbation indicates existence of a new family of
solutions which bifurcates from that of the even-dimensional MP black
holes.
We found instability in the gravitational scalar modes 
with several $\kappa$ satisfying $\kappa \geq 2$.
These perturbations break all of the symmetry in
\CPn~\cite{Dias:2010eu} and, thus, 
the branched solutions will have only $U(1)\times R_t$ isometry.
To confirm this picture by constructing 
non-linear solutions of the full geometry 
would be another direction of the future research.

We also found that the mass of the effective field on AdS$_2$, namely,
the eigenvalues of the angular part of the
perturbations are given by simple rational expressions. 
This property can be found also in the four and five-dimensional extreme black
holes and odd-dimensional MP black 
holes~\cite{Durkee:2010ea,Murata:2011my,Dias:2009ex,Amsel:2009ev}.
These results prompt us to conjecture that the eigenvalues for
the angular part of the perturbations are given by rational numbers for 
any vacuum near-horizon geometries with vanishing cosmological constant, 
as long as we assume the perturbations to be axisymmetric. 
Further studies on more general near-horizon geometries will be useful to 
falsify such a conjecture. 
One possible recipient for such an analysis is the 
near-horizon geometries of the odd-dimensional MP black holes with all but one 
angular momenta are equal, which have a cohomogeneity-1 near-horizon geometries.
The method used in this paper is directly applicable to these geometries, while it 
involves fifteen unknown variables in general. 

As a byproduct of our analysis,
we found that the expressions of 
the eigenvalues are essentially governed by the 
transformation property of the fields;
Other than a few exceptions, the 
eigenvalues of the scalar field coincide with
a part of those of the electromagnetic scalar 
modes, and the latter coincide with a part of 
those of the gravitational scalar mode.
We also found that the eigenvalues have simple organization described by
two integers $(C,\ell_0)$.
We find similar properties also for the vector modes, whose eigenvalues 
are described by Eq.~(\ref{lambda_gravVector}).
It would be interesting to clarify the mathematical and physical background for 
these properties.
It would also be nice if we can find microscopic interpretation
for them
based on the dual CFTs.

%


\subsection*{Acknowledgments}
\quad
We are grateful to Harvey Reall for useful comments on a draft of this paper.
N.T.~is supported in part by the DOE Grant DE-FG03-91ER40674.
K.M.~is supported by JSPS Grant-in-Aid for Scientific
Research No.24$\cdot$2337.

\appendix

\section{Background variables}
\label{App:basics}

In the appendices, we show the details of the method to solve the eigenvalue 
equations~(\ref{angularequations}).
In this App.~\ref{App:basics}, we introduce the basic objects which appear in the 
perturbation equations~(\ref{scalar})--(\ref{gravity}).
Using the formulae in this appendix, we show the explicit forms of the perturbation 
equations and show the analytic solutions for them when they exists in App.~\ref{App:calcEV}.
In the cases we cannot find analytic solutions, 
we solve equations numerically to obtain the eigenvalues.
In App.~\ref{App:Technical}, we summarize the numerical technique we employed.

Below, we work in the frame adapted to the spatial metric on the horizon~(\ref{horizonmetric})
given by
\begin{equation}
(\et)_\m = \Theta(d\theta)_\m~,
\qquad
(\ep)_\m = \Phi(d\phi + \A)_\m~,
\qquad
(e_\al)_\m = \Psi(\hat e_{\al})_\m~,
\label{frame}
\end{equation}
where $\hat e_\alpha$ are real orthonormal frame for \CPn.
We show the tetrad components of each tensor quantity below unless otherwise noted.
The spin connections 
$(\omega_{ab})_\mu \equiv (e_a)^\nu\na_\mu(e_b)_\nu$
for this frame are given by
\begin{equation}
\omega_{\theta\phi} = 
-\frac{(\log \Phi)'}{\Theta} e_\phi
,
\qquad
\omega_{\theta\alpha} 
=
-\frac{(\log \Psi)'}{\Theta}e_{\al}
,
\qquad
\omega_{\phi\alpha} = 
 \frac{\Phi}{\Psi^2}\J_{\al\be}e_{\be}
,
\qquad
\omega_{\al\be} = 
-\frac{\Phi}{\Psi^2}\J_{\al\be}\ep
+ \frac{1}{\Psi}\hat\omega_{\al\be}.
\end{equation}

Assuming the axisymmetry in our case, we have 
\begin{equation}
\begin{gathered}
k = \Omega \Phi e_\phi,
\qquad
k_Im^I = 0, 
\qquad
k_\nu k^\nu = \Omega^2\Phi^2,
\\
~~
dk = \Omega\left(
\frac{2\Phi}{\Theta} \left(\log \Phi\right)' e_\theta \wedge e_\phi
+ \frac{\Phi^2}{\Psi^2} J_{\alpha\beta} e_\alpha\wedge e_\beta
\right),
\qquad
dL^2 = \frac{2\Theta}{2n-1}\left(\log\Theta\right)'e_\theta.
\end{gathered}
\end{equation}

Using $\mathcal{R}_{ab}=d\omega_{ab}+\omega_{ac}\wedge \omega_{cb}$ and 
$de_a + \omega_{ab}\wedge e_b=0$, we find the curvature 2-forms are given by
\begin{align}
\mathcal{R}_{\theta\phi} &=
-\frac{1}{2\Theta^2}\left(
2(\log\Phi)'' + (\log\Phi)'\left(\log\frac\Phi\Theta\right)'
\right)
e_\theta\wedge e_\phi
- \frac{\Phi}{2\Psi^2\Theta}
\left(\log\frac{\Phi}{\Psi}\right)'
\J_{\alpha\be} e_\al \wedge e_\be~,
\label{curv_1}
\\
\mathcal{R}_{\theta\al} &=
-\frac{1}{2\Theta^2}\left(
2(\log\Psi)'' - (\log\Psi)'\left(\log\frac\Theta\alpha\right)'
\right) e_\theta\wedge e_\alpha
-\frac{\Phi}{2\Psi^2\Theta}\left(\log\frac\Phi\Psi\right)'
\J_{\alpha\beta} e_\phi\wedge e_\beta
~,
\label{curv_2}
\\
 \mathcal{R}_{\phi\alpha} 
&= 
\frac{\Phi}{2\Psi^2\Theta}
\left(\log\frac{\Phi}{\Psi}\right)' \J_{\alpha\beta}
\, e_\theta\wedge e_\beta
+
\left(
\frac{\Phi^2}{\Psi^4} - \frac{(\log\Phi)'(\log\Psi)'}{\Theta^2}
\right)e_\phi\wedge e_\alpha~,
\label{curv_3}
\\
\mathcal{R}_{\al\be} &=
- \frac{\Phi}{\Psi^2\Theta}
\left(\log\frac{\Phi}{\Psi}\right)' \J_{\al\be} \, e_\theta\wedge e_\phi
- \frac{{(\log\Psi)'}^2}{\Theta^2} e_\alpha\wedge e_\beta
+ \frac{1}{\Psi^2} \mathcal{\hat R}_{\al\be} 
-\frac{\Phi^2}{\Psi^4} \left(
\J_{\al\be}\J_{\ga\de} + \J_{\al[\ga|}\J_{\be|\de]}
\right) e_\ga\wedge e_\de~,
\label{curv_4}
\end{align}
where the Riemann and Ricci tensors of $\mathbb{CP}^{n-1}$ are given by
\begin{equation}
 \hat R_{\alpha\beta\gamma\delta} 
=
\de_{\alpha\gamma}\de_{\beta\delta} - \de_{\alpha\delta}\de_{\beta\gamma}
+ \J_{\alpha\gamma}\J_{\beta\delta}
- \J_{\alpha\delta}\J_{\beta\gamma}
+ 2\J_{\alpha\beta}\J_{\gamma\delta}~,
\qquad
\hat R_{\alpha\beta} = 2n\delta_{\alpha\beta}~.
\end{equation}

\section{Perturbation equations and eigenvalues}
\label{App:calcEV}

We sketch the procedure to construct the eigenvalue equations
for $\lambda_s$ in this appendix.
The procedure is parallel to that for
the odd-dimensional MP black holes with equal angular 
momenta~\cite{Durkee:2010ea}.
We show the analytic solutions and eigenvalues
for them when they exists, and numerical ones otherwise.
We defer a part of the derivations of the eigenvalue equations and most of the technical details to solve them
to App.~\ref{App:Technical}.
The calculations to obtain the tetrad components of the eigenvalue equations
mentioned below
are conducted partially
with the aid of the computer algebra system Cadabra~\cite{cadabra1,cadabra2}.

\subsection{Scalar field perturbations}
\label{App:scalar}

Since our interest is in the axisymmetric perturbations, we take 
the separation ansatz as 
\begin{equation}
\psi = \chi_0(t,r)f(\theta) \Y(x^\alpha),
\end{equation}
where $\Y$ is the scalar harmonics on \CPn defined with respect to the 
covariant derivative on \CPn, $D_\alpha$, as
\begin{equation}
\left(D^2 + \lS \right)\Y = 0,
\qquad
\lS = 4\kappa \left( \kappa + n - 1\right).
\qquad 
(\kappa=0,1,\ldots)
\label{ScalarHarmonics}
\end{equation}
Decomposing Eq.~(\ref{scalar}) into $\theta$ and \CPn parts and using 
Eq.~(\ref{ScalarHarmonics}), we find the eigenvalue equation to be
\begin{align}
\lambda_0 f\Y = 
 \mathcal{O}^{(0)} f\Y &= 
\frac{1}{2n-1}\left\{
-\6_\theta^2 f \Y
-\left( \log\bigl(\Theta\Phi\alpha^{2(n-1)}\bigr) \right)' \6_\theta f\Y
-\frac{\Theta^2}{\Psi^2} D^2 f\Y + \Theta^2 M^2 f\Y
\right\}
\no
&=
\frac{1}{2n-1}\left\{
-f'' -(2n-1)\cot\theta \,  f'
+  f_n(\theta) \left(
\frac{4\kappa(\kappa+n-1) }{2n\sin^2\theta}
+ 
\frac{(aM)^2 }{2n-1}
\right)  f
\right\} \Y
,
\label{O0}
\end{align}
where $f'\equiv df/d\theta$.
Note that the parameter $a$ in the metric~(\ref{metric}) only appears as a 
multiplicative factor of the mass term in Eq.~(\ref{O0}).
This property originates from the fact that 
$\lambda_0$ is dimensionless 
and thus only the dimensionless combination of the quantities, $aM$, may appear on 
the right-hand side.

When the scalar field is massless ($M=0$), Eq.~(\ref{O0}) has an exact solution 
given by
\begin{equation}
f(\theta) = C_1 P^\mu_\nu(\cos\theta) + C_2 (\sin\theta)^{1-n} Q^\mu_\nu(\cos\theta),
\end{equation}
where $P^\mu_\nu(x)$ and $Q^\mu_\nu(x)$ are the associated Legendre functions and
\begin{equation}
\mu = 2\kappa + n - 1,
\qquad
\nu = \frac12\left[
-1 + \sqrt{
\frac1n \left(2n-1\right)
\left\{
2 ( 2 \kappa + n - 1 )^2 - 2
+ n\left(4\lambda_0 + 3\right)
\right\}
}
\right].
\label{scalar_munu}
\end{equation}
The regularity of the solution
at $\theta = 0$
requires $C_2$ to be zero and $\nu$ to be an integer equal to or
larger than $\mu$.
This requirement gives the eigenvalue $\lambda_0$ shown in Eq.~(\ref{lambda0}).

When $M\neq 0$, we need to calculate $\lambda_0$ numerically.
From Eq.~(\ref{O0}), we find that a regular solution near $\theta=0$ behaves as
$f\sim\theta^{2\kappa}$. Behavior near $\theta = \pi$ is the same after 
$\theta\mapsto\pi-\theta$ due to the symmetry about $\theta=\pi/2$ of the 
near-horizon geometry (\ref{metric}).
Then, defining a new variable $\tilde f$ by $f = \tilde f \sin^{2\kappa}\theta $
and imposing the Neumann boundary condition at $\theta=0$ and $\pi$
to $\tilde f$,
we may obtain regular numerical solutions by the relaxation method.
We defer the further details of the numerical method to App.~\ref{App:Technical}.

We show the numerical value of $\lambda_0$ for $M=0$ and $\kappa=0$ 
along with the exact value given by Eq.~(\ref{lambda0}) in Table~\ref{Table:scalar_kappa0},
which shows the relative error of our numerical results from the exact results 
is maintained to be $\lesssim\mathcal{O}(10^{-5})$.
We also show the result for $aM=1$ and $\kappa=0$ in Table~\ref{Table:scalar_M1_kappa0}.
As we can see from these tables,  $\lambda_0$ tends to increase as $M$ increases.
As far as we have checked numerically, this property holds for any $M$, $\kappa$ and $n$.
If it holds in general, there is no instability for any $M^2\geq 0$.

\begin{table}[htbp]
\caption{
Numerical and exact values of
$\lambda_0$ for $M=0$ and $\kappa=0$.
}
\centering
\subtable[Numerical]{
{\scriptsize
\begin{tabular}{|c||c|c|c|c|c|} \hline
& n - 2 &&&&\\
mode\# & 0& 1& 2& 3& 4\\ \hline \hline
1 & 0.00001& 0.00001& 0.00001& 0.00001& 0.00001\\ \hline
2 & 1.33330& 1.20001& 1.14287& 1.11112& 1.09091\\ \hline
3 & 3.33334& 2.80000& 2.57143& 2.44445& 2.36364\\ \hline
4 & 5.99997& 4.80001& 4.28573& 4.00001& 3.81818\\ \hline
5 & 9.33333& 7.20001& 6.28572& 5.77778& 5.45455\\ \hline
6 & 13.33330& 10.00001& 8.57144& 7.77779& 7.27273\\ \hline
\end{tabular}
}
}
\subtable[Exact (Eq.~(\ref{lambda0}))]{
{\scriptsize
\begin{tabular}{|c||c|c|c|c|c|} \hline
& n - 2 &&&&\\
mode\# & 0& 1& 2& 3& 4\\ \hline \hline
1 & 0.00000& 0.00000& 0.00000& 0.00000& 0.00000\\ \hline
2 & 1.33333& 1.20000& 1.14286& 1.11111& 1.09091\\ \hline
3 & 3.33333& 2.80000& 2.57143& 2.44444& 2.36364\\ \hline
4 & 6.00000& 4.80000& 4.28571& 4.00000& 3.81818\\ \hline
5 & 9.33333& 7.20000& 6.28571& 5.77778& 5.45455\\ \hline
6 & 13.33333& 10.00000& 8.57143& 7.77778& 7.27273\\ \hline
\end{tabular}
}
}
\label{Table:scalar_kappa0}
\end{table}

\begin{table}[htbp]
\caption{
Numerical value of $\lambda_0$ for $a M=1$ and $\kappa=0$.
}
\centering
\begin{tabular}{|c||c|c|c|c|c|} \hline
& n - 2 &&&&\\
mode\# & 0& 1& 2& 3& 4\\ \hline \hline
1 & 0.176311& 0.068203& 0.036151& 0.022391& 0.015231\\ \hline
2 & 1.586020& 1.306218& 1.202036& 1.149000& 1.117304\\ \hline
3 & 3.600436& 2.920479& 2.641832& 2.491085& 2.396961\\ \hline
4 & 6.271692& 4.927228& 4.362393& 4.052046& 3.856095\\ \hline
5 & 9.607142& 7.330994& 6.366260& 5.833399& 5.495654\\ \hline
6 & 13.608308& 10.133272& 8.654517& 7.835879& 7.316149\\ \hline
\end{tabular}
\label{Table:scalar_M1_kappa0}
\end{table}

\subsection{Electromagnetic perturbations}
\label{Sec:Maxwell}

Next, we discuss the perturbations of the Maxwell field on the near-horizon geometries.
For the analysis, it is useful to decompose the perturbation variables $Y_\mu$ into 
the scalar and vector modes according to the transformation properties with respect to \CPn.
We show the derivations of the eigenvalue equations and results mode by mode below.

\subsubsection{Electromagnetic vector modes}


The electromagnetic vector modes are characterized by
\begin{equation}
Y_i=0, \qquad D^{\pm\alpha} Y_\alpha =0,
\qquad 
(i=\theta,\phi)
\label{def_EMvector}
\end{equation}
where $D^{\pm}$ is the covariant derivative on \CPn
projected to the $\mp i$ eigenspaces of the complex structure for \CPn,
$\J = \frac12 \J_{\alpha}\hat e_\alpha \hat e_\beta$,
defined as
\begin{equation}
D^{\pm}_{\alpha} \equiv \proj_\alpha^{\pm\beta} D_\beta,
\qquad
\proj^\pm_{\alpha\beta}
=
\frac12 \left(
\hat g_{\alpha\beta} \pm i \J_{\alpha\beta}
\right).
\label{proj}
\end{equation}
We may parametrize this component of the perturbations as
$Y_\alpha = f(\theta)\Y_\alpha$, where $\Y_\alpha$ is the divergence-free vector 
harmonic satisfying
\begin{equation}
D^{\pm\alpha}\Y_\alpha = 0,
\qquad
\Y^\pm_{\alpha\beta} = 
\frac{-1}{\sqrt{\lambda^V_\kappa}} D^\pm{\!\!\!\!}_{(\alpha}\Y_{\beta)},
\qquad
\left( D^2 + \lV \right) \Y_\alpha = 0,
\qquad
\lV = 4\kappa(\kappa+2) + 2n \left(2\kappa + 3\right).
\label{VectorHarmonics}
\end{equation}
For these modes, we find $i$ components of 
Eq.~(\ref{Maxwell}) becomes trivial, and \CPn components 
reduce to an equation of $f(\theta)$ given by
\begin{multline}
 -n(2n-1) \lambda_1 f(\theta) = 
f''(\theta )
+
\frac{ (2n-1)\left\{(2n+1)\cos^2\theta-1\right\} }
{\tan\theta \left\{(2n-1)\cos^2\theta + 1\right\} }
f'(\theta ) 
\\
+
\left\{
\frac{
(2n-1)^2 \cos^2\theta +4n - 1
}{\left\{(2n-1)\cos^2\theta +1\right\}^2 }
-
\frac{(2 \kappa +3) (2 n+2 \kappa +1) }{\sin^2\theta}
\right.
\\
\left.
+
\frac{2 (2 n-1) \left((n+2) \kappa +n+\kappa^2\right)}{n}
\right\}
f(\theta ).
\label{Eq_EMvector}
\end{multline}
This equation has an analytic solution given by
\begin{equation}
f(\theta) = 
C_1 P^\mu_\nu(\cos\theta) + C_2 (\sin\theta)^{1-n} Q^\mu_\nu(\cos\theta),
\label{EMvector_sol}
\end{equation}
where 
{\small
\begin{equation}
\mu = 2\kappa + n +2,
\qquad
\nu =
\frac12\left[
-1 + \sqrt{
\frac1n \left\{
4 n^3+4 n^2 (4 \kappa +2 \lambda_1+5)+n (8 \kappa  (2 \kappa +3)-4
   \lambda_1-7)-8 \kappa  (\kappa +2)
\right\}
}
\right].
\label{EMvector_munu}
\end{equation}
}%
For the solution to be regular at $\theta=0$, 
$C_2$ should be set to zero and 
$\nu$ should be an integer equal to or larger than $\mu$.
This requirement fixes $\lambda_1$ as given in Eq.~(\ref{lambda_EVvector}).

\subsubsection{Electromagnetic scalar modes}

Next, we discuss the electromagnetic scalar modes for which $i$ components of 
the perturbations are turned on. We may expand the scalar mode perturbations as 
\begin{equation}
Y_i = f_i(\theta) \Y,
\qquad
Y_\alpha = 
  g^+(\theta) \Y^+_\alpha
+ g^-(\theta) \Y^-_\alpha,
\label{def_EMscalar}
\end{equation}
where 
$\Y^\pm_\alpha$ are the scalar-derived vector eigenfunctions defined by 
\begin{equation}
\Y_\alpha^\pm \equiv -\frac{D^\pm_\alpha \Y}{\sqrt{\lambda_\kappa^S}},
\qquad
\Y^{\pm\pm}_{\alpha\beta} \equiv
D^\pm_{(\alpha}\Y^\pm_{\beta)},
\qquad
\Y^{+-}_{\alpha\beta}\equiv
  D^+_{(\alpha}\Y_{\beta)}^-
+ D^-_{(\alpha}\Y_{\beta)}^+
- \frac{\sqrt{\lambda_\kappa^S}}{2(n-1)} \hat g_{\alpha\beta} \Y.
\label{scalarDerivedTensors}
\end{equation}
We further assume that $Y_\mu$ are real-valued, 
and introduce the real quantities by 
\begin{equation}
 g^\pm = g^R \pm ig^I,
\qquad
\Y^\pm_\alpha = \Y^R_\alpha \pm i \Y^I_\alpha.
\label{EMscalar_gdef}
\end{equation}
As a result, we find coupled ODEs for $(f_\theta, f_\phi, g^R, g^I)$
as the eigenvalue equations when $\kappa>0$.
For $\kappa=0$, $\Y$ becomes a constant and 
the variables reduce to $(f_\theta,f_\phi)$.
We solve those equations using the numerical technique demonstrated in 
App.~\ref{App:Technical}.

In Table~\ref{Table:EMscalar}, we show an example of the numerically-obtained 
eigenvalues $\lambda_1$.
As far as we have examined,
our numerical result is consistent with a hypothesis that 
the eigenvalues are expressed in general by Eqs.~(\ref{lambda_EMscalar}),
(\ref{lambda_EMscalar_list}) and (\ref{lambda_EMscalar_list_k0}).

\begin{table}[htbp]
\caption{
Numerical value of $\lambda_1$ of the electromagnetic scalar modes
for $\kappa = 4$ and $n=2,\ldots,6$.
}
\label{Table:EMscalar}
\centering
\begin{tabular}{|c||c|c|c|c|c|} \hline
 $\kappa = 4$ & $n - 2$ &&&&\\ \cline{1-1}
mode\# & 0& 1& 2& 3& 4\\ \hline \hline
1 & 4.000000& 1.600000& 0.857143& 0.533333& 0.363636\\ \hline 
2 & 9.333331& 4.799999& 3.142856& 2.311111& 1.818181\\ \hline 
3 & 9.999997& 5.599998& 3.999999& 3.199999& 2.727272\\ \hline 
4 & 9.999997& 5.599998& 3.999999& 3.199999& 2.727272\\ \hline 
5 & 15.999992& 9.199996& 6.571425& 5.199997& 4.363634\\ \hline 
6 & 16.666658& 9.999995& 7.428567& 6.088885& 5.272724\\ \hline 
7 & 16.666656& 9.999994& 7.428566& 6.088885& 5.272723\\ \hline 
8 & 16.666658& 9.999995& 7.428568& 6.088885& 5.272724\\ \hline 
9 & 23.333315& 13.999989& 10.285706& 8.311104& 7.090903\\ \hline 
10 & 23.999980& 14.799987& 11.142846& 9.199990& 7.999991\\ \hline 
11 & 23.999983& 14.799990& 11.142849& 9.199993& 7.999994\\ \hline 
12 & 23.999981& 14.799988& 11.142847& 9.199991& 7.999992\\ \hline 
\end{tabular}
\end{table}

\subsection{Gravitational perturbations}
\label{Sec:Gravity}

Finally, we study the gravitational perturbations.
The procedure is parallel to that for the electromagnetic perturbations, 
while the gravitational perturbations involve ten unknown variables in general.
We show the procedures and results for each of tensor, vector and scalar modes below.

\subsubsection{Gravitational tensor modes}
\label{App:gravtensor}

We start from the simplest components, which are the gravitational 
tensor modes 
defined by
\begin{equation}
 Y_{ij}=Y_{i\alpha} = 0,
\qquad
\hat g^{\al\be}Y_{\al\be}=0,
\qquad
 D^{\pm \al}Y_{\al\be}=0.
\end{equation}
Under these assumptions, only the \CPn components of 
Eq.~(\ref{gravity}) remain nontrivial.
We introduce a decomposition given by
\begin{equation}
Y_{\alpha\beta}=f(\theta)\mathbb{Y}_{\alpha\beta},
\label{def_gravtensor}
\end{equation}
and
we also decompose $\mathbb{Y}_{\alpha\beta}$ into hermitian and 
anti-hermitian parts by
$ \left(\J \mathbb{Y}\J\right)_{\alpha\beta} = \sigma \mathbb{Y}_{\alpha\beta}$,
where $\sigma=-1\,(+1)$ for the hermitian (anti-hermitian) modes.

To deal with the $D^2Y_{\alpha\beta}$ terms in the equations, we need to take 
$\mathbb{Y}_{\alpha\beta}$ to be an eigenstate of the generalized Lichnerowicz operator
on $\mathbb{CP}^{n-1}$, that is,
\begin{equation}
\lT
\mathbb{Y}_{\alpha\beta}
=
\left( \Delta_\text{L}^\A \mathbb{Y} \right)_{\alpha\beta}
\equiv
-D^2\mathbb{Y}_{\alpha\beta} - 2\hat R_{\alpha\gamma\beta\delta}\mathbb{Y}^{\gamma\delta}
+ 4n \mathbb{Y}_{\alpha\beta}
=
-D^2\mathbb{Y}_{\alpha\beta} +2(2n+1)\mathbb{Y}_{\alpha\beta}
+6 \left(\J \mathbb{Y}\J\right)_{\alpha\beta}.
\end{equation}
For $n=2$, the tensor harmonics do not exist on $\mathbb{CP}^1=S^2$.
For $n\geq 3$, the harmonics exist and
the eigenvalues of $\Delta_\text{L}^\A $ 
are given by~\cite{Kunduri:2006qa}
\begin{equation}
\lT
= 
4 \kappa \left( \kappa + n - 1 \right)
+ 4\left(n-1+\sigma\right).
\end{equation}
This allows us to rewrite $D^2 Y_{\alpha\beta}$ in terms of $Y_{\alpha\beta}$.
Using the above equations, 
the eigenvalue equation (\ref{gravity}) is rewritten as
\begin{multline}
-(2n-1)\lambda_2 \, f(\theta) = 
f''(\theta)
+ 
(2n-1)\left[
\cot \theta 
-\frac{2\sin 2 \theta }{(2 n-1) \cos^2 \theta +1}
\right]
f'(\theta)
\\
+ 
2
\left[
2 n \left(
\frac{2}{(2 n-1) \cos^2 \theta +1}
+\kappa -1
\right)
- \frac{2 \kappa  (n+\kappa -1)}{\sin^2\theta}
\right.
\\
\left.
+\frac{-\kappa ^2+\kappa -\sigma +1}{n}+\kappa  (2 \kappa
   -3)+2 \sigma -3
\right]
f(\theta).
\label{eq_gravtensor}
\end{multline}
This equation has an exact solution given by
\begin{equation}
 f(\theta)=  C_1 P^\mu_\nu (\cos\theta) 
+ C_2 (\sin\theta)^{1-n} Q^\mu_\nu(\cos\theta),
\end{equation}
where
\begin{equation}
 \mu = 2\kappa+n-1,
\qquad
\nu=
\frac{1}{2} \left(
-1 + 
\sqrt{
\frac{ 2 n-1 }{n}
\left\{
2\left( 2\kappa+n-1 \right)^2
-10 + n\left(4\lambda_2+3\right) + 8\sigma
\right\}
}
\right).
\end{equation}
Setting
$C_2=0$ and $\nu\geq \mu$ to be an integer
for the sake of the regularity at $\theta=0$,
we find $\lambda_2$ to be given by Eq.~(\ref{lambda_gravtensor}).

\subsubsection{Gravitational vector modes}
\label{App:gravvector}

The gravitational vector modes are composed of divergence free vectors 
$Y_{i\alpha}$ and traceless $Y_{\alpha\beta}$ satisfying
\begin{equation}
 Y_{ij} = 0,
\qquad
D^{\pm\alpha} Y_{i\alpha} = 0,
\qquad
Y_{\alpha}^{~\alpha}=0.
\end{equation}
We expand the perturbations as
\begin{equation}
 Y_{i\alpha} = g_i(\theta) \Y_\alpha,
\qquad
Y_{\alpha\beta} = 
h^+(\theta) \Y^+_{\alpha\beta} + h^-(\theta) \Y^-_{\alpha\beta}
\equiv
Y^+_{\alpha\beta} + Y^-_{\alpha\beta},
\label{def_gravvector}
\end{equation}
where $\Y_{\alpha}$ is a divergence-free vector harmonics defined by 
Eq.~(\ref{VectorHarmonics}). We further decompose $\Y_\alpha$ into the 
eigenvectors of $\J_{\alpha\beta}$ by 
$\J_\alpha^{~\beta}\Y_\beta = -i\varepsilon \Y_\alpha$ with 
$\varepsilon = \pm 1$.

Since $\Y_\alpha$ are complex, the variables $(g_i, (h^+ + h^-)/2,(h^+-h^-)/2i)$
are not real-valued.
To rewrite the variables in terms of real quantities,
we further decompose the variables as%
\begin{equation}
  Y_{i\alpha} = g_i^+(\theta) \tY^+_\alpha + g_i^-(\theta) \tY^-_\alpha,
\qquad
Y_{\alpha\beta} = 
  h^{++}(\theta) \tY^{++}_{\alpha\beta} 
+ h^{-+}(\theta) \tY^{-+}_{\alpha\beta}
+ h^{+-}(\theta) \tY^{+-}_{\alpha\beta} 
+ h^{--}(\theta) \tY^{--}_{\alpha\beta},
\end{equation}
where $\tY_\alpha^\pm$ are the vector harmonics
corresponding to $\varepsilon = \pm 1$
(that is, $\J_\alpha^{~\beta}\tY^\pm_\beta = \mp i \tY_\alpha^\pm$)
and 
\begin{equation}
\tY^{\pm_1\pm_2}_{\alpha\beta} \equiv
\frac{-1}{\sqrt{\lambda^V_\kappa}} D^{\pm_1}_{(\alpha}\tY^{\pm_2}_{\beta)}
~.
\end{equation}
Then, replacing the quantities by
\begin{equation}
\begin{aligned}
 g^\pm_i &= g^R_i \pm i g^I_i,
& \quad
 h^{\pm_1\pm_2} &= 
\bigl(h^{RR} \pm_2 i h^{RI}\bigr) 
\pm_1 i \bigl( h^{IR} \pm_2 i h^{II}\bigr),
\\
\tY^\pm_i &= \tY^R_i \pm i \tY^I_i,
& \quad
 \tY^{\pm_1\pm_2}_{\alpha\beta} &= 
\bigl(\tY^{RR}_{\alpha\beta} \pm_2 i \tY^{RI}_{\alpha\beta}\bigr) 
\pm_1 i \bigl( \tY^{IR}_{\alpha\beta} \pm_2 i \tY^{II}_{\alpha\beta}\bigr),
\end{aligned}
\label{gravvector_vardef}
\end{equation}
we obtain real-valued eigenvalue equations in terms of 
$(g^R_\theta, g^I_\theta, g^R_\phi, g^I_\phi, h^{RR}, h^{RI}, h^{IR}, h^{II})$.

Due to the self-adjointness of the operators $\mathcal{O}^{(s)}$
with respect to suitable inner products,
$\lambda_2$ is guaranteed to be real numbers~\cite{Durkee:2010ea}.
This fact implies that the $\epsilon=\pm 1$ modes share the same eigenvalue $\lambda_2$.

We show the numerically-obtained eigenvalues in Table~\ref{Table:gravVector}.
Note that this result contains the eigenvalues for both $\epsilon=1$ 
and $\epsilon=-1$ modes, which are equal to each other.
This numerical result suggest that 
there are four species of the eigenvalues, 
other than the multiplicity coming from
$\epsilon=\pm 1$ modes, 
and the expression of the eigenvalues for general $n$ and $\kappa$ is
given by Eqs.~(\ref{lambda_gravVector}) and (\ref{lambda_gravVector_list}).
These eigenvalues are all positive, and hence no instability is implied.

\begin{table}[htbp]
\caption{$\lambda_2$ of gravitational vector modes
for $n=5$ and $\kappa=0,\ldots,3$.
}
\label{Table:gravVector}
\centering
\subtable{
{ \small
\begin{tabular}{|c||c|c|c|c|} \hline
 $n = 5$ & $\kappa$ &&&\\ \cline{1-1}
mode\# & 0& 1& 2& 3\\ \hline \hline
1 & 0.4444& 0.5778& 0.8000& 1.1111\\ \hline 
2 & 0.4444& 0.5778& 0.8000& 1.1111\\ \hline 
3 & 0.8889& 1.4667& 2.1333& 2.8889\\ \hline 
4 & 0.8889& 1.4667& 2.1333& 2.8889\\ \hline 
5 & 2.0000& 2.5778& 3.2444& 4.0000\\ \hline 
6 & 2.0000& 2.5778& 3.2444& 4.0000\\ \hline 
7 & 2.0000& 2.5778& 3.2444& 4.0000\\ \hline 
8 & 2.0000& 2.5778& 3.2444& 4.0000\\ \hline 
9 & 2.6667& 3.6889& 4.8000& 6.0000\\ \hline 
10 & 2.6667& 3.6889& 4.8000& 6.0000\\ \hline 
11 & 3.7778& 4.8000& 5.9111& 7.1111\\ \hline 
12 & 3.7778& 4.8000& 5.9111& 7.1111\\ \hline 
\end{tabular}
}
}
\subtable{
{ \small
\begin{tabular}{|c||c|c|c|c|} \hline
 $n = 5$ & $\kappa$ &&&\\ \cline{1-1}
mode\# & 0& 1& 2& 3\\ \hline \hline
13 & 3.7778& 4.8000& 5.9111& 7.1111\\ \hline 
14 & 3.7778& 4.8000& 5.9111& 7.1111\\ \hline 
15 & 3.7778& 4.8000& 5.9111& 7.1111\\ \hline 
16 & 3.7778& 4.8000& 5.9111& 7.1111\\ \hline 
17 & 4.6667& 6.1333& 7.6889& 9.3333\\ \hline 
18 & 4.6667& 6.1333& 7.6889& 9.3333\\ \hline 
19 & 5.7778& 7.2444& 8.8000& 10.4444\\ \hline 
20 & 5.7778& 7.2444& 8.8000& 10.4444\\ \hline 
21 & 5.7778& 7.2444& 8.8000& 10.4444\\ \hline 
22 & 5.7778& 7.2444& 8.8000& 10.4444\\ \hline 
23 & 5.7778& 7.2444& 8.8000& 10.4444\\ \hline 
24 & 5.7778& 7.2444& 8.8000& 10.4444\\ \hline 
\end{tabular}
}
}
\end{table}

\subsubsection{Gravitational scalar modes}

Finally, we comment on the gravitational scalar modes, for which 
the analysis procedure is similar to the previous examples while it is more involved.

We expand the metric perturbations using the scalar harmonics $\Y$ as
\begin{equation}
\begin{aligned}
Y_{ij} &= f_{ij}(\theta) \Y ,
\\
Y_{i\alpha} &= g_i^+(\theta) \Y^+_\alpha + g_i^-(\theta) \Y^-_\alpha,
\\
Y_{\alpha\beta} &=
-\frac{1}{\sqrt{\lambda_\kappa^S}}\left(
  h^{++}(\theta) \Y^{++}_{\alpha\beta}
+ h^{--}(\theta) \Y^{--}_{\alpha\beta}
+ h^{+-}(\theta) \Y^{+-}_{\alpha\beta}
\right)
-\frac{1}{2(n-1)}
\bigl(f_{\theta\theta}(\theta)+f_{\phi\phi}(\theta)\bigr)
\delta_{\alpha\beta}\Y,
\end{aligned}
\label{def_gravscalar}
\end{equation}
where
$\Y^\pm_\alpha$ and $\Y^{\pm\pm}_{\alpha\beta}$ are scalar-derived 
vector/tensor eigenfunctions defined by Eq.~(\ref{scalarDerivedTensors}).
To construct the real-valued equations, 
we redefine the variables and mode functions as 
\begin{equation}
g^\pm_i \equiv g^R_i \pm i g^I_i,
\qquad
h^{\pm\pm} \equiv h^R \pm i h^I,
\qquad
\Y^\pm_\alpha \equiv \Y^R_\alpha \pm i \Y^I_\alpha,
\qquad
\Y^{\pm\pm}_{\alpha\beta} \equiv \Y^R_{\alpha\beta} \pm i \Y^I_{\alpha\beta}.
\label{gravscalar_vardef}
\end{equation}
As a result, we obtain coupled ODEs of 
$(f_{\theta\theta}, f_{\theta\phi}, f_{\phi\phi}, 
g^R_\theta, g^I_\theta, g^R_\phi, g^I_\phi, h^R, h^I, h^{+-})$ in a general case
($\kappa>1$ and $n>2$) as the eigenvalue equations to solve.
Depending on the values of $\kappa$ and $n$, some of the unknown variables drop
out as follows (see also Ref.~\cite{Durkee:2010ea}):
\begin{itemize}
\item $\kappa=0$:
$\mathbb{Y}$ becomes a constant 
 and there is no scalar-derived vector nor tensor in this case.
We have only $(f_{ij})$ as the unknown variables.

\item $\kappa=1$ \& $n=2$:
All of $\mathbb{Y}^{\pm\pm}_{\alpha\beta}$ and $\mathbb{Y}^{+-}_{\alpha\beta}$
vanish in this case, and we have $\bigl(f_{ij},g^{R,I}_i\bigr)$ as 
the unknown variables. 

\item $\kappa=1$ \& $n>2$:
$\mathbb{Y}^{\pm\pm}_{\alpha\beta}$ vanish in this case, and we have $\bigl(f_{ij},g^{R,I}_i,h^{+-}\bigr)$ as 
the unknown variables.

\item $\kappa>1$ \& $n=2$:
$\mathbb{Y}^{+-}_{\alpha\beta}$ vanishes in this case.
As a result, we have $\bigl(f_{ij},g^{R,I}_i,h^{\pm\pm}\bigr)$ as the unknown variables.
\end{itemize}

The numerically-obtained eigenvalues for $\kappa>1$ and $n>2$
suggest that they are described by 
Eqs.~(\ref{lambda_gravScalar_simple}) and (\ref{lambda_gravScalar_simple2}).
We show some examples of the numerically obtained eigenvalues $\lambda_2$ 
in Table~\ref{Table:gravscalar}.
For the special cases mentioned above,
we find that the eigenvalues are given by Eq.~(\ref{lambda_gravScalar_simple})
in any cases, and only some of the integer sets $(C, \ell_0)$ are 
modified. We list the integer sets for each case
in Eqs.(\ref{lambda_gravscalar_list_k0})--(\ref{lambda_gravscalar_list_n2}).
In short, some of the eigenvalues listed in Eq.~(\ref{lambda_gravScalar_simple2})
vanishes, and also $\ell_0$ is shifted in some of the remaining ones.

\begin{table}[htbp]
\caption{$\lambda_2$ of gravitational scalar modes
for $n=2,5$ and $\kappa=0,\ldots,4$.
}
\label{Table:gravscalar}
\centering
\subtable{
{ \footnotesize
\begin{tabular}{|c||c|c|c|c|c|} \hline
 $n = 2$ & $\kappa$ &&&&\\ \cline{1-1}
mode\# & 0& 1& 2& 3& 4\\ \hline \hline
1 & 3.333& 1.333& -0.667& -0.667& -0.000\\ \hline 
2 & 4.000& 2.000& 0.667& 2.000& 4.000\\ \hline 
3 & 5.333& 2.000& 2.000& 3.333& 5.333\\ \hline 
4 & 6.000& 3.333& 2.000& 3.333& 5.333\\ \hline 
5 & 6.667& 4.000& 3.333& 6.000& 9.333\\ \hline 
6 & 8.000& 4.667& 4.000& 6.667& 10.000\\ \hline 
7 & 9.333& 4.667& 4.000& 6.667& 10.000\\ \hline 
8 & 10.000& 4.667& 5.333& 8.000& 11.333\\ \hline 
9 & 11.333& 6.000& 5.333& 8.000& 11.333\\ \hline 
10 & 13.333& 6.000& 5.333& 8.000& 11.333\\ \hline 
11 & 14.000& 7.333& 7.333& 11.333& 16.000\\ \hline 
12 & 15.333& 8.000& 8.000& 12.000& 16.667\\ \hline 
13 & 18.000& 8.000& 8.000& 12.000& 16.667\\ \hline 
14 & 18.667& 8.000& 8.000& 12.000& 16.667\\ \hline 
15 & 20.000& 9.333& 9.333& 13.333& 18.000\\ \hline 
16 & 23.333& 9.333& 9.333& 13.333& 18.000\\ \hline 
17 & 24.000& 9.333& 9.333& 13.333& 18.000\\ \hline 
18 & 25.333& 11.333& 9.333& 13.333& 18.000\\ \hline 
19 & 29.333& 12.000& 12.000& 17.333& 23.333\\ \hline 
20 & 30.000& 12.000& 12.667& 18.000& 24.000\\ \hline 
\end{tabular}
}
}
\subtable{
{ \footnotesize
\begin{tabular}{|c||c|c|c|c|c|} \hline
 $n = 5$ & $\kappa$ &&&&\\ \cline{1-1}
mode\# & 0& 1& 2& 3& 4\\ \hline \hline
1 & 2.444& 0.444& -0.356& -0.622& -0.800\\ \hline 
2 & 3.333& 1.333& 0.089& 0.267& 0.533\\ \hline 
3 & 4.000& 1.333& 0.978& 1.378& 1.644\\ \hline 
4 & 4.444& 2.000& 1.200& 1.378& 1.644\\ \hline 
5 & 4.889& 2.444& 1.200& 1.600& 2.311\\ \hline 
6 & 5.778& 2.444& 1.867& 2.489& 3.200\\ \hline 
7 & 6.000& 2.889& 1.867& 2.489& 3.200\\ \hline 
8 & 6.667& 2.889& 2.978& 3.600& 4.311\\ \hline 
9 & 7.778& 2.889& 2.978& 3.600& 4.311\\ \hline 
10 & 7.778& 3.778& 2.978& 3.600& 4.311\\ \hline 
11 & 8.667& 4.000& 2.978& 3.600& 4.311\\ \hline 
12 & 9.778& 4.000& 2.978& 4.044& 5.200\\ \hline 
13 & 10.000& 4.000& 3.867& 4.933& 6.089\\ \hline 
14 & 10.889& 4.667& 3.867& 4.933& 6.089\\ \hline 
15 & 12.000& 4.667& 3.867& 4.933& 6.089\\ \hline 
16 & 12.444& 4.667& 4.978& 6.044& 7.200\\ \hline 
17 & 13.333& 5.778& 4.978& 6.044& 7.200\\ \hline 
18 & 14.444& 5.778& 4.978& 6.044& 7.200\\ \hline 
19 & 15.111& 5.778& 4.978& 6.044& 7.200\\ \hline 
20 & 16.000& 5.778& 4.978& 6.044& 7.200\\ \hline 
\end{tabular}
}
}
\end{table}

\section{Technical details}
\label{App:Technical}

We summarize the numerical technique to solve the eigenvalue equations
in this appendix.
This technique is essentially the same as that used in Ref.~\cite{Murata:2011my},
which studied the conformal weights for the general 
near-horizon geometries in five dimensions.

\subsection{Perturbation equations and boundary conditions
for electromagnetic scalar modes}
\label{Sec:BC_EM}

We explain the numerical technique to find the eigenvalues taking 
the electromagnetic scalar modes for example.
%
%
For the electromagnetic scalar modes defined by Eq.~(\ref{def_EMscalar}),
we obtain a system of coupled ODEs for 
the unknown variables $(f_\theta, f_\phi, g^R, g^I)$ 
if $\kappa>0$, where these variables are defined by Eqs.~(\ref{def_EMscalar}) and
(\ref{EMscalar_gdef}).
If $\kappa=0$, the scalar harmonics $\Y$ becomes a constant, and the unknown 
variables reduce to $(f_\theta, f_\phi)$.

Below, we take the $\kappa>0$ case as an example, and explain the boundary 
conditions to be imposed on the unknown variables.
The resultant eigenvalue equations may be expressed as
\begin{equation}
\left(
\6_\theta^2  + A(\theta) \6_\theta + B(\theta)
\right)\vec v = -(2n-1) \lambda_1 \vec v,
\label{eqAB}
\end{equation}
where 
$
\vec v \equiv
\left( f_\theta, f_\phi, g^R, g^I \right)^T
$,
and $A(\theta)$ and $B(\theta)$ 
are $4\times 4 $ coefficient matrices which behave near $\theta=0$ as
\begin{equation}
A = \frac{2n-1}{\theta} E + \mathcal{O}\bigl(\theta^0\bigr),
\qquad
B = \frac{1}{\theta^2} B_{-2} + \mathcal{O}\bigl(\theta^{-1}\bigr),
\end{equation}
where $E$ is the unit matrix.
To find out the fall-off behavior of the solution at $\theta=0$, we need to 
diagonalize the coefficient matrix $B_{-2}$
by introducing new variables $\vec {\tilde v}$ by $\vec v = \Xi \vec{\tilde v}$,
where $\Xi$ is the matrix composed of the eigenvectors of $B_{-2}$.
The equations in terms of these new variables are 
\begin{equation}
\left(
\6_\theta^2  + \tilde A(\theta) \6_\theta + \tilde B(\theta)
\right)
\vec {\tilde v}
= -(2n-1) \lambda_1 \vec {\tilde v},
\label{eqABtilde}
\end{equation}
where $\tilde A = \frac{2n-1}{\theta} E + \mathcal{O}(\theta^0)$ and 
$\tilde B$ behaves for $\theta \to 0$ as
\begin{equation}
\tilde B = \frac{1}{\theta^2} \tilde B_{-2} + \mathcal{O}\bigl(\theta^{-1}\bigr),
\qquad
\tilde B_{-2} = \text{Diag}\left[
\epsilon_-, \epsilon_-, \epsilon_+, \epsilon_+
\right],
\qquad
\epsilon_\pm \equiv
- \left(2\kappa \pm 1\right)\left\{
2(n-1) + 2\kappa  \pm 1
\right\}.
\end{equation}
Solving Eq.~(\ref{eqABtilde}) near $\theta=0$, 
we find the asymptotic behavior of the solution near $\theta=0$ to be given by
\begin{equation}
\vec {\tilde v}_i \propto \theta ^{p_i},
\qquad
p_i = 
\begin{cases}
2\kappa -1 & (i = 1, 2) \\
2\kappa +1 & (i = 3,4) \\
\end{cases}
~,
\end{equation}
where we chose the decaying solutions from the two linearly-independent solutions 
so that the solutions are regular at $\theta=0$.
Based on this observation, we introduce new variables $\vec{\hat v}$ by
$\vec {\tilde v} = \Pi \vec {\hat v}$, where $\Pi$ is the diagonal matrix whose 
components are given by $\Pi_{ii} = \sin^{p_i} \theta$.
For these new variables $\vec {\hat v}$, we find the eigenvalue equations become
\begin{equation}
\left(
\6_\theta^2  + \hat A(\theta) \6_\theta + \hat B(\theta)
\right)
\vec {\hat v}
= -(2n-1) \lambda_1 \vec {\hat v},
\label{eqABhat}
\end{equation}
where $\hat A = \mathcal{O}(\theta^{-1})$ and $\hat B=\mathcal{O}(\theta^0)$.
This equation has a regular singularity at $\theta=0$, and we need to impose the 
Neumann boundary condition there, 
namely, $\vec{\hat v}= \vec{\hat v}_0 + \mathcal{O}(\theta^2)$, to maintain the 
regularity of the solutions.

Since the perturbation equations (\ref{Maxwell}) become symmetric with respect to 
$\theta=\pi/2$ for the near-horizon metric~(\ref{metric}), 
we may solve Eq.~(\ref{eqABhat}) by imposing the Neumann boundary 
condition at both $\theta=0$ and $\theta=\pi$.
Alternatively, we separate the variables into the even/odd modes with respect to 
$\theta=\pi/2$ as follows.
In Eq.~(\ref{eqAB}), we find the parity with respect to $\theta=\pi/2$
of the coefficient matrix components are given as
\begin{equation}
A=
\begin{array}{r}
{\scriptstyle i=1 \big\{} \\
{\scriptstyle i=2,3,4 \big\{} 
\end{array}
\!\!\!\!
\overset{\vphantom{1}}{
\left(
\begin{array}{c|c}
\smash{\overbrace{\vphantom{\frac11}(\text{odd})}^{i=1}}
& 
\smash{\overbrace{\vphantom{\frac11}(\text{even})}^{i=2,3,4}}
\\ \hline
(\text{even})&(\text{odd})
\end{array}
\right)
},
\qquad
B=
\left(
\begin{array}{c|c}
(\text{even}) & (\text{odd}) \\ \hline
(\text{odd}) & (\text{even})
\end{array}
\right).
\end{equation}
This property originates from that the original equations are symmetric with 
respect to $\theta=\pi/2$, and also that the variable $v_1 = f_\theta$ is 
an odd function with respect to $\theta=\pi/2$, since it is proportional to the
$e_\theta$ component of the perturbation variables, while $v_{2,3,4}$ are not.
This property implies that an eigenvector take a form either of
\begin{equation}
\vec v = 
\begin{array}{r}
{\scriptstyle i=1 \big\{} \\
{\scriptstyle i=2,3,4 \big\{} 
\end{array}
\!\!\!\!
\left(
\begin{array}{c}
v_i^{(o)}
\\ 
v_i^{(e)}
\end{array}
\right),
\qquad
\vec v = 
\left(
\begin{array}{c}
v_i^{(e)}
\\ 
v_i^{(o)}
\end{array}
\right),
\label{EVoe}
\end{equation}
where $v_i^{(o)}$ and $v_i^{(e)}$ are odd and even functions
with respect to $\theta=\pi/2$, respectively.
Since both $A$ and $B$ are regular at $\theta=\pi/2$,
we may impose the Dirichlet boundary condition on $v_1$ to set it zero
and the Neumann boundary condition on $v_{2,3,4}$ to obtain solutions 
corresponding to the left one in Eq.~(\ref{EVoe}).
Solutions corresponding to the right one in Eq.~(\ref{EVoe}) can be obtained by 
switching the boundary condition types.
By decomposing into the even/odd modes, we may reduce the calculation region
$0<\theta < \pi$ to the half, and also reduce the number of modes needs to be 
subtracted at each time step in the relaxation method, which will be introduced 
in the next section, to the half. It helps increasing the precision of the 
numerical results in a shorter calculation time.

In practice, 
the even/odd decomposition is more simply implemented by introducing
new variables 
$\vec{\check v}$ by $\vec v = \Upsilon \vec{\check v}$,
where $\Upsilon$ is a diagonal matrix whose components are equal to%
\footnote{
The function $w=\cos\bigl(\frac{\pi}{2}(1-\cos\theta)\bigr)$ is chosen so that 
$w$ and $w/(\theta-\pi/2)$ are sufficiently close to constants
near $\theta=0$ and $\theta=\pi/2$, respectively.
For example, simpler choices such as $w=\cos\theta$ or $w=\cos(2\theta^2/\pi)$
give rise to $\mathcal{O}(\theta^{-2})$ or 
$\mathcal{O}\bigl((\theta-\frac{\pi}{2})^{-1}\bigr)$ terms in 
the components of the matrix $\hat B$ in some cases.
Such terms in $\check B$ become obstacles for our scheme.
}
$\cos\bigl(\frac{\pi}{2}(1-\cos\theta)\bigr)$
for $i=1$ ($i=2,3,4$)
and $1$ for $i=2,3,4$ ($i=1$) if we are to find the solutions of the left
(right) type in Eq.~(\ref{EVoe}). 
In terms of $\vec {\check v}$, the eigenvalue equations become
\begin{equation}
\left(
\6_\theta^2  + \check A(\theta) \6_\theta + \check B(\theta)
\right)
\vec {\check v}
= -(2n-1) \lambda_1 \vec {\check v},
\label{eqABcheck}
\end{equation}
where $\check A=\mathcal{O}\bigl((\frac\pi2-\theta)^{-1}\bigr)$ and 
$\check B(\theta)=\mathcal{O}\bigl((\frac\pi2-\theta)^0\bigr)$,
and we may impose the Neumann boundary condition on all the components of 
$\vec{\check v}$ to find the solutions of the left (right) type in Eq.~(\ref{EVoe}).
Following the procedures in the previous paragraphs after introduction of $\Upsilon$, 
we define and use the variables given by
$\vec {\hat v} = (\Upsilon\Xi\Pi)^{-1} \vec v$,
and solve Eq.~(\ref{eqABhat}) imposing the Neumann boundary condition at
$\theta=0$ and $\pi/2$.

For $\kappa=0$, the unknown variables reduce to 
$\vec v = (f_\theta, f_\phi)$ as we mentioned previously. We find 
$B_{-2}=\text{Diag}[-(2n-1), -(2n-1)]$ without the diagonalization, 
and we find that the regular solutions behave as $v_i \propto \theta^0$.
Thus, we may solve the eigenvalue equations using the original variables $v_i$, 
or alternatively introducing only the even/odd decomposition defined by 
$\Upsilon=\text{Diag}(w,1)$ 
or 
$\Upsilon=\text{Diag}(1,w)$ with $w\equiv\cos\bigl(\frac{\pi}{2}(1-\cos\theta)\bigr)$.

\subsection{Numerical implementations}
\label{Sec:Numerics}

We solve the eigenvalue equation~(\ref{eqABhat}) numerically as follows.
We employ the relaxation method, which was used also in Ref.~\cite{Murata:2011my} 
to treat similar problems.

Firstly, we introduce a diffusion equation given by
\begin{equation}
\6_\tau \vec {\hat v}(\tau, \theta) = M(\theta) \vec{\hat v},
\qquad
M(\theta) \equiv
\6_\theta^2 + \hat A(\theta) \6_\theta + B(\theta),
\label{diffusionequation}
\end{equation}
whose solution behaves as
\begin{equation}
\vec {\hat v}(\tau, \theta) 
= e^{\tau M} \vec {\hat v}_\text{init} 
\to e^{\tau \lambda_*}\vec{\hat v}_*, 
\end{equation}
where $\lambda_*$ is the largest eigenvalue of $M$
and $\vec{\hat v}_*$ is the corresponding eigenvector.
It implies that we may obtain the largest eigenvalue by following the time 
evolution described by Eq.~(\ref{diffusionequation}) for a sufficiently long time.
The smaller eigenvalues can be obtained successively by the same calculations
if we project out the eigenvectors for larger eigenvalues at each step of 
the time evolution.

To make the equation amenable to numerics,
we need to discretize the diffusion equation (\ref{diffusionequation}) 
on the grids given by $\tau = \tau_n \equiv n \Delta \tau$ $(n=0,1,\ldots)$
and $\theta = \theta_i\equiv i \Delta \theta$ $(i=0,1,\ldots,N,~ 
\Delta\theta\equiv \pi/ 2N)$.
We do it implicitly as, for $1\leq i \leq N-1$,
\begin{equation}
\frac{
\vec{\hat v}_i^n - \vec{\hat v}_i^{n-1}
}{\Delta\tau}
=
\frac{
\vec{\hat v}_{i+1}^n -2 \vec{\hat v}_i^n + \vec{\hat v}_{i-1}^n
}{\Delta\theta^2}
+ \hat A_i \frac{
\vec{\hat v}_{i+1}^n - \vec{\hat v}_{i-1}^n
}{2\Delta\theta}
+ \hat B_i \vec{\hat v}_{i}^n.
\label{discretizedequation}
\end{equation}
An advantage of the implicit scheme is that 
the time evolution is stable for any $\Delta \tau$, 
and we may take it large to speed up the numerical calculations.
Eq.~(\ref{discretizedequation}) may be transformed into
\begin{align}
\vec{\hat v}_{i}^{n-1} &= 
\left(
-\frac{\Delta\tau}{\Delta\theta^2}E 
+\frac{\Delta\tau}{2\Delta\theta}\hat A_i
\right) \vec{\hat v}_{i-1}^n
+\left(
\Bigl(
1+\frac{2\Delta\tau}{\Delta\theta^2} 
\Bigr) E - \Delta\tau \hat B_i
\right)\hat v_i^n
+
\left(
-\frac{\Delta\tau}{\Delta\theta^2}E 
-\frac{\Delta\tau}{2\Delta\theta}\hat A_i
\right) \vec{\hat v}_{i+1}^n
\notag \\
&\equiv
  a_i \vec {\hat v}_{i-1}^n
+ b_i \vec {\hat v}_{i}^n
+ c_i \vec {\hat v}_{i+1}^n.
\label{eqabc}
\end{align}
We need to impose the Neumann boundary condition at $\theta=0$ and 
$\theta=\pi/2$. At $\theta=0$, it implies 
$\vec{\hat v} \simeq \vec {\hat v}_0 + \frac12 \vec {\hat v}''_0 \theta^2$,
and Eq.~(\ref{diffusionequation}) at $\theta=0$ may be expressed as
\begin{equation}
\6_\tau \vec{\hat v} = 
\left(
E + \hat{\hat A}_0
\right)\6^2_\theta \vec{\hat v}
+ \hat B_0 \vec{\hat v},
\end{equation}
where $\hat{\hat A}_0\equiv \lim_{\theta\to 0}\theta \hat A(\theta)$.
This equation is discretized to give an equation for $i=0$ as
\begin{gather}
\frac{\vec{\hat v}_0^n - \vec{\hat v}_0^{n-1}}{\Delta\tau}
=
\frac{2}{\Delta\theta^2}
\left( E + \hat{\hat A}_0 \right)
\left(
\vec{\hat v}_1^n - \vec{\hat v}_0^n
\right) 
+ \hat B_0 \vec{\hat v}_0^n
\no
\Leftrightarrow \qquad
 v_0^{n-1}
=
\left(
\Bigl(1+\frac{2\Delta\tau}{\Delta\theta^2}\Bigr)E
+ \frac{2\Delta\tau}{\Delta\theta^2}\hat{\hat A}_0
- \Delta\tau \hat B_0
\right) \vec{\hat v}_0^n
- \frac{2\Delta\tau}{\Delta\theta^2}\left(
E + \hat{\hat A}_0
\right) \vec{\hat v}_1^n
\equiv
   b_0 \vec{\hat v}_0^n
+  c_0 \vec{\hat v}_1^n.
\end{gather}
Following the same procedure for $\theta=\pi/2$
, we have the equation at $i=N$ as
\begin{equation}
 v_N^{n-1}
=
- \frac{2\Delta\tau}{\Delta\theta^2}\left(
E + \hat{\hat A}_N
\right) \vec{\hat v}_{N-1}^n
+
\left(
\Bigl(1+\frac{2\Delta\tau}{\Delta\theta^2}\Bigr)E
+ \frac{2\Delta\tau}{\Delta\theta^2}\hat{\hat A}_N
- \Delta\tau \hat B_N
\right) \vec{\hat v}_N^n
\equiv
   a_N \vec{\hat v}_{N-1}^n
+  b_N \vec{\hat v}_N^n,
\end{equation}
where 
$\hat{\hat A}_N \equiv \lim_{\theta\to\frac\pi2}\bigl(\theta-\frac\pi2\bigr)\hat A(\theta)$.

The above equations are summarized as follows:
\begin{equation}
\left(
\begin{array}{l}
\hat v_0^{n-1}
\\
\hat v_1^{n-1}
\\
~~~\vdots
\\
\hat v_{N-1}^{n-1}
\\
\hat v_N^{n-1}
\end{array}
\right)
=
\left(
\begin{array}{
ccccc
}
b_0 & c_0 & & \\
a_1 & b_1 & c_1 & &\\
         & \ddots  & \ddots& \ddots& \\
 & &a_{N-1} &b_{N-1} & c_{N-1}\\
 & & &a_N &b_N \\
\end{array}
\right)
\left(
\begin{array}{l}
\hat v_0^n
\\
\hat v_1^n
\\
~~~\vdots
\\
\hat v_{N-1}^n
\\
\hat v_N^n
\end{array}
\right).
\label{eqmatrix}
\end{equation}
This equation is readily solved 
by the forward and backward substitution to obtain $\vec{\hat v}^n$ from $\vec{\hat v}^{n-1}$.
To obtain the full spectrum, we need to conduct two calculations for 
the two $\Upsilon$ matrices, each of which corresponds to the each mode in Eq.~(\ref{EVoe}).

Applying this method to Eq.~(\ref{eqABhat}) for the electromagnetic scalar modes, 
we obtain the results summarized in Table~\ref{Table:EMscalar}
and also as 
Eq.~(\ref{lambda_EMscalar}) supplemented with Eq.~(\ref{lambda_EMscalar_list}).
Similarly, we may obtain the results for the $\kappa=0$ case, and the results are 
summarized as Eq.~(\ref{lambda_EMscalar_list_k0}).
In this calculation, we took $N=4000$ and $\Delta \tau = 10^{-2}$,
and truncated the time evolution 
once the condition
$|\lambda^{-1}\frac{d\lambda}{d\tau}|\lesssim 10^{-8}$ is achieved.
After the truncation, we progressed the time evolution one more step using smaller
$\Delta \tau = 10^{-8}$ to obtain the correct value of $\lambda_1$.

\subsection{Gravitational vector modes}

The gravitational vector modes have the following properties.
The unknown variables are 
\begin{equation}
\vec v =
\bigl(
g^R_\theta, g^I_\theta, 
g^R_\phi, g^I_\phi, 
h^{RR}, h^{RI}, h^{IR}, h^{II}
\bigr)^T,
\label{v_gV}
\end{equation}
whose definitions are found in Eq.~(\ref{gravvector_vardef}).
Following the procedure of Sec.~\ref{Sec:BC_EM} in this case, we find the 
diagonalized coefficient matrix $\tilde B_{-2}$ is given by
\begin{equation}
\tilde B_{-2} = \text{Diag}\left[
\epsilon_-, \epsilon_-, \epsilon_-, \epsilon_-, 
\epsilon_+, \epsilon_+, \epsilon_+, \epsilon_+
\right],
\qquad
\epsilon_- \equiv -4(\kappa+2)(n+\kappa),
\qquad
\epsilon_+ \equiv -4(\kappa+2)(n+\kappa+1).
\end{equation}
Solving the equation corresponding to Eq.~(\ref{eqABtilde}), we find the 
asymptotic behaviors for $\theta\to 0$ of the decaying solutions to be
\begin{equation}
\vec{\tilde v}\sim \left(
\theta^{p_-},\theta^{p_-},
\theta^{p_-},\theta^{p_-},
\theta^{p_+},\theta^{p_+},
\theta^{p_+},\theta^{p_+}
\right)^T,
\qquad
  p_- = 2(\kappa+1), 
\qquad
 p_+ = 2(\kappa+2).
\end{equation}
Then, we define and use new variables $\vec{\hat v}\equiv \Pi^{-1}\vec{\tilde v}$,
where $\Pi$ is a diagonal matrix whose components are
given by $\Pi_{ii}=\sin^{p_-}\theta$ for 
$i=1,\ldots,4$ and $\Pi_{ii}=\sin^{p_+}\theta$ for $i=5,\ldots,8$.
Following the procedure in Secs.~\ref{Sec:BC_EM} and \ref{Sec:Numerics}, 
we obtain the eigenvalues summarized in Table~\ref{Table:gravVector} and 
also as Eqs.~(\ref{lambda_gravVector}) and (\ref{lambda_gravVector_list}).

\subsection{Gravitational scalar modes}

The gravitational scalar modes have the following properties.
For a general case with $\kappa>1$ and $n>2$,
the unknown variables are given by
\begin{equation}
\vec v = 
\bigl(f_{\theta\theta}, f_{\theta\phi}, f_{\phi\phi}, 
g^R_\theta, g^I_\theta, g^R_\phi, g^I_\phi, h^R, h^I, h^{+-}\bigr)^T,
\end{equation}
whose definitions are found in Eq.~(\ref{gravscalar_vardef}).
$\tilde B_{-2}$ for this case becomes
\begin{equation}
\tilde B_{-2}
= 
\text{Diag}
\left[ 
\epsilon_-,\epsilon_-,\epsilon_-,
\epsilon_0,\epsilon_0,\epsilon_0,\epsilon_0,
\epsilon_+,\epsilon_+,\epsilon_+
\right],
\qquad
\epsilon_\delta
= 
-4 \left(\kappa + \delta\right) \left(\kappa + n -1 + \delta \right),
\label{B2_gravscalar}
\end{equation}
which give rise to the asymptotic behavior described as
\begin{equation}
\vec {\tilde v} \sim 
\left(
\theta^{p_-},\theta^{p_-},\theta^{p_-},
\theta^{p_0},\theta^{p_0},\theta^{p_0},\theta^{p_0},
\theta^{p_+},\theta^{p_+},\theta^{p_+}
\right)^T,
\qquad
p_\delta \equiv 2\left(\kappa + \delta\right).
\end{equation}
We define and use the matrix $\Pi$ using $p_\delta$ in this equation.
The numerical results of the eigenvalue $\lambda_2$ are summarized as 
Eq.~(\ref{lambda_EMscalar}) with Eq.~(\ref{lambda_EMscalar_list}).

Below, we comment on the special cases for the gravitational scalar modes.
\begin{itemize}
\item $\kappa=0$:

$\mathbb{Y}=\text{constant}$ in this case,
and the unknown variables are given by 
$\vec v = (f_{\theta\theta}, f_{\theta\phi}, f_{\phi\phi})$.
Without introducing the diagonalization, we find 
$B_{-2} = \text{Diag}[-4n,-4n,-4n]$, from which we find that 
the regular solutions behave as $v_i \propto \theta^2$.
We define and use $\Xi=E$ and $\Pi$ such that $\Pi_{ii}= \sin^2\theta$.

\item $\kappa=1$ \& $n=2$:

All of $\mathbb{Y}^{\pm\pm}_{\alpha\beta}$ and $\mathbb{Y}^{+-}_{\alpha\beta}$ vanish, and 
as a result we have only 
$(
f_{\theta\theta},f_{\theta\phi},f_{\phi\phi},
g^{R}_\theta, g^{I}_\theta, 
g^{R}_\phi,g^{I}_\phi
)$ 
as the free variables in this case.
It follows
$\tilde B_{-2}=\text{Diag}[
0,-8,-8,-8,-24,-24,-24
]$, from which we find 
$\tilde v_i \propto \theta^{p_i}$ with 
$p_{1}=0$, $p_{2,3,4}=2$ and $p_{5,6,7}=4$.

\item $\kappa=1$ \& $n>2$:

$\mathbb{Y}^{\pm\pm}_{\alpha\beta}$ vanishes in this case, and 
we have 
$\vec v=(
f_{\theta\theta},f_{\theta\phi},f_{\phi\phi},
g^{R}_\theta, g^{I}_\theta, 
g^{R}_\phi,g^{I}_\phi,
h^{+-}
)^T$ 
as the free variables. We find
$\tilde B_{-2}=\text{Diag}[0,-4n,-4n,-4n,-8(n+1),-8(n+1),-8(n+1)]$,
which results in 
$\tilde v_i\propto \theta^{p_i}$ with $p_{1}=0$, $p_{2,3,4}=2$ and $p_{5,6,7}=4$.

\item $\kappa>1$ \& $n=2$:

$\mathbb{Y}^{+-}_{\alpha\beta}$ vanishes in this case, and
as a result we have 
$(
f_{\theta\theta},f_{\theta\phi},f_{\phi\phi},
g^{R}_\theta, g^{I}_\theta, 
g^{R}_\phi,g^{I}_\phi,
h^{++},h^{--} 
)$ 
as the free variables.
We find
$\tilde B_{-2}=\text{Diag}[
\epsilon_-,\epsilon_-,\epsilon_-,
\epsilon_0,\epsilon_0,\epsilon_0,
\epsilon_+,\epsilon_+,\epsilon_+
]$, where $\epsilon_\delta$ are those in Eq.~(\ref{B2_gravscalar})
with $n=2$ plugged in.
The asymptotic behavior of the regular solutions in this case is given by
$\tilde v_i \propto \theta^{p_i}$ with $p_{1,2,3}=2(\kappa-1)$, 
$p_{4,5,6}=2\kappa$ and $p_{7,8,9}=2(\kappa+1)$.
\end{itemize}

The numerical results of the eigenvalues $\lambda_2$ for these cases are 
summarized by Eqs.~(\ref{lambda_gravscalar_list_k0})--(\ref{lambda_gravscalar_list_n2}).

\end{document}